\documentclass{aa}  

\usepackage{graphicx}
\usepackage{epstopdf}
\usepackage{booktabs}
\usepackage{color}
\usepackage{txfonts}
\usepackage[colorlinks = true,
            linkcolor = blue,
            urlcolor  = blue,
            citecolor = blue,
            anchorcolor = blue]{hyperref}

\def\lesssim{\mathrel{\hbox{\rlap{\hbox{\lower4pt\hbox{$\sim$}}}\hbox{$<$}}}}
\def\gtrsim{\mathrel{\hbox{\rlap{\hbox{\lower4pt\hbox{$\sim$}}}\hbox{$>$}}}}

\def\msol{\hbox{\kern 0.20em $M_\odot$}}
\def\lsol{\hbox{\kern 0.20em $L_\odot$}}
\def\rsol{\hbox{\kern 0.20em $R_\odot$}}
\def\sr{\hbox{\kern 0.20em sr}}
\def\srmu{\hbox{\kern 0.20em sr$^{-1}$}}

\def\g{\hbox{\kern 0.20em g}}
\def\gmu{\hbox{\kern 0.20em g$^{-1}$}}
\def\kg{\hbox{\kern 0.20em kg}}
\def\pc{\hbox{\kern 0.20em pc}}

\def\mum{\hbox{\kern 0.20em $\mu$m}}
\def\mumd{\hbox{\kern 0.20em $\mu$m$^{-2}$}}
\def\cm{\hbox{\kern 0.20em cm}}
\def\m{\hbox{\kern 0.20em m}}
\def\km{\hbox{\kern 0.20em km}}
\def\nm{\hbox{\kern 0.20em nm}}

\def\s{\hbox{\kern 0.20em s}}
\def\h{\hbox{\kern 0.20em h}}
\def\sec{\hbox{\kern 0.20em sec}}
\def\min{\hbox {\kern 0.20em min}}
\def\smu{\hbox{\kern 0.20em s$^{-1}$}}
\def\smd{\hbox{\kern 0.20em s$^{-2}$}}
\def\an{\hbox{\kern 0.20em an}}
\def\anmu{\hbox{\kern 0.20em an$^{-1}$}}
\def\deg{\hbox{\kern 0.20em $^{\rm o}$}}
\def\yr{\hbox{\kern 0.20em yr}}
\def\yrmu{\hbox{\kern 0.20em yr$^{-1}$}}
\def\Myr{\hbox{\kern 0.20em Myr}}
\def\Mymu{\hbox{\kern 0.20em Myr$^{-1}$}}
\def\K{\hbox{\kern 0.20em K}}
\def\pcmu{\hbox{\kern 0.20em pc$^{-1}$}}
\def\pcmd{\hbox{\kern 0.20em pc$^{-2}$}}
\def\pcmt{\hbox{\kern 0.20em pc$^{-3}$}}
\def\kms{\hbox{\kern 0.20em km\kern 0.20em s$^{-1}$}}
\def\kmpd{\hbox{\kern 0.20em km$^{2}$}}
\def\kpc{\hbox{\kern 0.20em kpc}}
\def\cms{\hbox{\kern 0.20em cm\kern 0.20em s$^{-1}$}}
\def\erg{\hbox{\kern 0.20em erg}}
\def\ergs{\hbox{\kern 0.20em erg}}
\def\cmpd{\hbox{\kern 0.20em cm$^2$}}
\def\cmmd{\hbox{\kern 0.20em cm$^{-2}$}}
\def\cmms{\hbox{\kern 0.20em cm$^{-6}$}}
\def\cmpt{\hbox{\kern 0.20em cm$^3$}}
\def\cmmt{\hbox{\kern 0.20em cm$^{-3}$}}
\def\mpd{\hbox{\kern 0.20em m$^2$}}
\def\mmd{\hbox{\kern 0.20em m$^{-2}$}}
\def\mpt{\hbox{\kern 0.20em m$^3$}}
\def\mmt{\hbox{\kern 0.20em m$^{-3}$}}
\def\mujy{\hbox{\kern 0.20em $\mu$Jy}}
\def\mjy{\hbox{\kern 0.20em mJy}}
\def\Mj{\hbox{\kern 0.20em MJy}}
\def\jy{\hbox{\kern 0.20em Jy}}
\def\ghz{\hbox{\kern 0.20em GHz}}
\def\srmd{\hbox{\kern 0.20em sr$^{-1}$}}
\def \kms{km~$\rm{s}^{-1}$}

\def \mum{$\mu$m}

\def\G{\hbox{\kern 0.20em G}}

\def\h13cop{\hbox{H$^{13}$CO$^{+}$}}

\def\S+{\hbox{S{\small II}}}

\begin{document} 

\title{Infrared signature of active massive black holes\\in nearby dwarf galaxies\thanks{Tables 2 and 3 are only available in electronic form at the CDS via anonymous ftp to cdsarc.u-strasbg.fr (130.79.128.5) or via http://cdsweb.u-strasbg.fr/cgi-bin/qcat?J/A+A/}}

\author{
Francine R. Marleau\inst{\ref{inst1}}
\and Dominic Clancy\inst{\ref{inst1}}
\and Rebecca Habas\inst{\ref{inst1}}
\and Matteo Bianconi\inst{\ref{inst1},}\inst{\ref{inst2}}
}

\institute{
Institute of Astro and Particle Physics, University of Innsbruck, Technikerstra{\ss}e 
25, 6020 Innsbruck, Austria\\ 
e-mail: \href{mailto:Francine.Marleau@uibk.ac.at}{Francine.Marleau@uibk.ac.at};
\href{mailto:Dominic.Clancy@uibk.ac.at}{Dominic.Clancy@uibk.ac.at};
\href{mailto:Rebecca.Habas@uibk.ac.at}{Rebecca.Habas@uibk.ac.at}
\label{inst1}
\and
Astrophysics and Space Research Group, School of Physics and Astronomy, 
University of Birmingham,\\Edgbaston, Birmingham B15 2TT United Kingdom,
e-mail: \href{mailto:mbianconi@star.sr.bham.ac.uk}{mbianconi@star.sr.bham.ac.uk}
\label{inst2}
}

\date{}

\titlerunning{CMBH in Nearby Dwarf Galaxies}

\authorrunning{Marleau et al.} 



  \abstract
  {We investigate the possible presence of active galactic nuclei (AGN) 
  in dwarf galaxies and other nearby galaxies to identify candidates 
  for follow-up confirmation and dynamical mass measurements.}
  {We identify candidate active central massive black holes (CMBH) using their 
  mid-infrared emission, verify their nature using existing 
  catalogues and optical line emission diagnostics, and study the 
  relationship between their mass and the mass of their host galaxy.}
  {We use the Wide-field Infrared Survey Explorer (WISE)
  All-Sky Release Source Catalog and examine the infrared colours of a
  sample of dwarf galaxies and other nearby galaxies in order to 
  identify both unobscured and obscured candidate AGN
  by applying the infrared colour diagnostic. Stellar
  masses of galaxies are obtained using a combination of three
  independent methods. Black hole masses are estimated using the
  bolometric luminosity of the AGN candidates and computed for three
  cases of the bolometric-to-Eddington luminosity ratio.}
  {We identify 303 candidate AGN, of which 276 were subsequently 
  found to have been independently identified as AGN via other methods. The
  remaining 9\% require follow-up observations for confirmation. The
  activity is detected in galaxies with stellar masses from $\sim 10^6$ to
  $10^9$ M$_{\odot}$; assuming the candidates are AGN, the black
  hole masses are estimated to be $\sim 10^3 - 10^6$ M$_{\odot}$,
  adopting $L_{bol} = 0.1 L_{Edd}$. The black hole masses probed are
  several orders of magnitude smaller than previously reported for
  centrally located massive black holes. We examine the stellar mass
  versus black hole mass relationship in this low galaxy mass
  regime. We find that it is consistent with the existing relation
  extending linearly (in log-log space) into the lower mass regime.}
  {These findings suggest that
  CMBH are present in low-mass galaxies and in
  the Local Universe, and provide new impetus for follow-up dynamical
  studies of quiescent black holes in local dwarf galaxies.}


   \maketitle


\section{Introduction}
\label{sec:introduction}

Following pioneering work in the late 1960s
\citep{zeldovich65,salpeter64,lyndenbell69,bardeen70,lyndenbell71}, a
paradigm has gradually emerged in which central massive black holes
(CMBH) have come to be regarded as an integral component of ``most, if
not all, massive galaxies,'' intimately linked to their formation and
evolution
\citep{merritt01,bender03,ho00,peterson08,peterson10,heckman11,schawinski12,kormendy13}.
However, despite a great deal of research and progress on many fronts
towards a better understanding of this paradigm \citep[see e.g.]
[for a recent review]{merloni15}, essentially all of the
fundamental questions concerning the formation, growth, and host
co-evolution of CMBH remain unanswered. One such question, which is
the focal point of this article, is whether CMBH are
generic to all galaxies. It is interesting to note that while it is
often stated in the literature that black holes are to be found at the
centres of most galaxies, until recently there has been insufficient
evidence to support this claim, even for the case of massive galaxies
not possessing a significant bulge component, let alone the general
population of galaxies, most of which are not massive. In other words, it is not
generally known whether CMBH are only a generic feature of certain
galaxy types, present only in galaxies above a certain mass threshold,
or subject to a combination of both of these restrictions.

The question of CMBH genericity is strongly related to the issues of
formation, growth and galaxy co-evolution. Its relationship with the
subject of galaxy formation can be understood by considering the
current observational constraints on high-redshift
quasars \citep[e.g.][]{fan01,mortlock11,wu15,matsuoka16}, which imply
that CMBH were present at very early times ($z > 7$) and therefore 
must have formed either concurrently with their host galaxies or
prior to them. While these constraints clearly highlight the close
connection between the formation of CMBH and their hosts, they also
conversely imply that galaxies which do not host a CMBH will likely
have undergone distinct formation processes. Similarly, in order to
understand the link between CMBH genericity and evolution, one need
only consider the evidence of the significant effects that active
galactic nuclei (AGN) have on their hosts' evolutions
\citep{silk98,king03,fabian12,silk13,ishibashi14} in order to realise
that galaxies which do not host CMBH may have undergone a markedly
different evolution from those that do.

It is apparent from these simple observations that the question of
CMBH genericity has important implications for the theory and
modelling of galaxy and large-scale structure evolution \citep[see
  e.g.][for recent reviews]{benson10,silk13}. Indeed, the
importance of modelling AGN feedback has been understood for some time
\citep{springel05,dimatteo05,mcnamara07} and various approaches are
now incorporated as standard in simulations of galaxy and structure
evolution
\citep{sijacki07,dimatteo08,khalatyan08,booth09,power11,degraf12,newton13}.
Feedback from AGN has also been shown to play a significant role up
to the largest scales of structure \citet{haider2016} \citep[see
  also][for a comparison of models]{wurster13}. In general, it is
expected that theories and models which assume that black holes are
generic will differ significantly in their predictions from those that
do not. For example, while most models and simulations at present
include the effects of AGN feedback from massive galaxies, including
the effects of AGN feedback from low-mass galaxies may account for the
observed low baryon fraction in Milky Way-type galaxies at the present
epoch \citep[see e.g.][]{peirani12}.

While the broad question of the genericity of CMBH remains open,
recent findings have been strongly supportive of CMBH being generic
for the subset of massive galaxies, i.e.\ for galaxies with stellar
masses above $\sim10^9$ M$_{\odot}$. In particular, in a recent work
\citet{marleau13} studied a sample of 15~991~galaxies and found the
fraction of galaxies containing a CMBH to be approximately the same
for each morphological type. This study was itself prompted by the
recent discovery of CMBH in galaxies lacking any substantial
spheroidal component \citep{reines11,simmons13}. Prior to this there
was only strong evidence for the generic presence of CMBH in galaxies
with a significant bulge component
\citep{kormendy95,magorrian98,ferrarese00,gebhardt00,kormendy13}. While
evidence continues to accrue to firmly establish the generic existence
of CMBH within massive galaxies, attention has now begun to shift to
the question of whether CMBH are also generic in the low-mass regime,
i.e.\ in galaxies with stellar masses below $\sim10^9$ M$_{\odot}$,
or in other words, whether CMBH are generic to dwarf galaxies.

Until very recently it was not even known whether any dwarf galaxies
contained CMBH, let alone whether they were generic. However, during
the last few years results in this area have been rapid. It is now
known that a significant number of galaxies with masses
$\sim10^9-10^{10}$ M$_{\odot}$, i.e.\ at the high end of the dwarf
galaxy mass regime and the low end of the massive galaxy regime, likely 
contain active CMBH.  In particular, \citet{moran14},
\citet{reines13}, \citet{reines14}, \citet{barth08}, \citet{greene07},
\citet{greene04}, \citet{dong12} and \citet{dong07} have detected the
optical signatures of mainly unobscured (type~1) AGN in 768 galaxies
with total stellar masses in the range $\sim10^8-10^{11}$ M$_{\odot}$,
while \citet{marleau13} have identified the first mid-infrared signatures
of both obscured and unobscured candidate AGN in 73 dwarf galaxies with masses 
in the range $\sim10^7-10^9$ M$_{\odot}$.

Even though these findings are encouraging, they barely probe the
dwarf galaxy mass range, and it is also the case that the black hole
masses derived from the optical spectra may be subject to large
systematic uncertainties.  Additionally, one would ideally also like
to have dynamically derived masses. At the present time, there is only
one dynamical measurement of a black hole in a galaxy classified as a
dwarf galaxy, i.e.\ with a mass below $3 \times 10^9$
M$_{\odot}$, namely the CMBH in NGC~4395, which has a mass of $4
\times 10^5$ M$_{\odot}$ \citep{denBrok15}. Additionally,
\citet{seth14} found a black hole of mass $2.1 \times 10^7$
M$_{\odot}$ in the ultra-compact dwarf galaxy M60-UCD1 of total stellar
mass $1.2 \times 10^8$ M$_{\odot}$, though this is believed to be a
tidally stripped galaxy whose progenitor had a mass of $\sim10^{10}$
M$_{\odot}$, which would then be consistent with the known black hole
mass to host total stellar mass scaling relation \citep{marleau13}.

As dwarf galaxies exhibit a number of properties that differentiate
them from other galaxies \citep[see review in][]{mateo98} and are
known to have different evolutionary histories from massive galaxies,
this could be a consequence of them not having CMBH or of having a
different relationship with their CMBH. If it is the case that dwarf
galaxies generically contain CMBH, it is therefore not necessarily the
case that what has been learned about the relationships between CMBH
and their hosts in the massive regime, as exhibited in their scaling
relations, will necessarily hold for dwarf galaxies. However, the
extension of current scaling relations suggests that dwarf galaxies
should host an intermediate mass CMBH, if they host one at all. The
low surface brightness of dwarfs necessitates studying them in the
nearby universe; consequently, this population of galaxies provides
the opportunity to discover very nearby, low-mass CMBH ($<
10^6$ M$_{\odot}$), which would be ideal for follow-up dynamical
studies.

Motivated by our recent infrared (IR) study of active massive black
holes in galaxies of all morphological types \citep{marleau13} and
wide range of stellar masses, we have undertaken a census in the very
nearby Universe, specifically targeting low-mass dwarf systems that
should be ideal hosts for CMBH in the mass range of intermediate mass 
black holes (IMBH; $M_{BH} \sim
10^3 - 10^6$ M$_{\odot}$). Additionally, we are interested in
detecting AGN in nearby galaxies regardless of their mass so that we
can identify targets for future dynamical mass measurement.  The
structure of our paper is as follows. In Section~\ref{sec:sample}, we
describe our dwarfs and other nearby galaxies sample. In
Sections~\ref{sec:IRcolor} and \ref{sec:IRAGN}, this sample is matched
to an infrared catalogue and the AGN candidates are identified using
the infrared colour diagnostic. The detection of these AGN are verified
using existing catalogues and optical line emission diagnostic in
Section~\ref{sec:verification} and their distance distribution is
presented in Section~\ref{sec:distance}. In Sections~\ref{sec:mstar}
and \ref{sec:agnfraction}, we derive stellar masses for the host
galaxies of our AGN candidates and compute the AGN fraction as a
function of stellar mass based on our AGN selection method. In
Sections~\ref{sec:bhmass} and \ref{sec:relation}, we evaluate the black
hole masses for our AGN candidates, present the black hole mass versus
stellar mass scaling relation, and compare our black hole mass
estimates to those derived using other methods. In
Section~\ref{sec:discussion}, we discuss and summarize our results. We believe
that by firmly establishing the presence of active IMBHs not only at
the low end of the galaxy mass function but also in the nearby
Universe, it will be possible to follow-up their quiet counterparts
with dynamical studies and hence provide further support for their
existence.

\begin{figure*}
\centerline{
\includegraphics[width=170pt,height=170pt,angle=0]{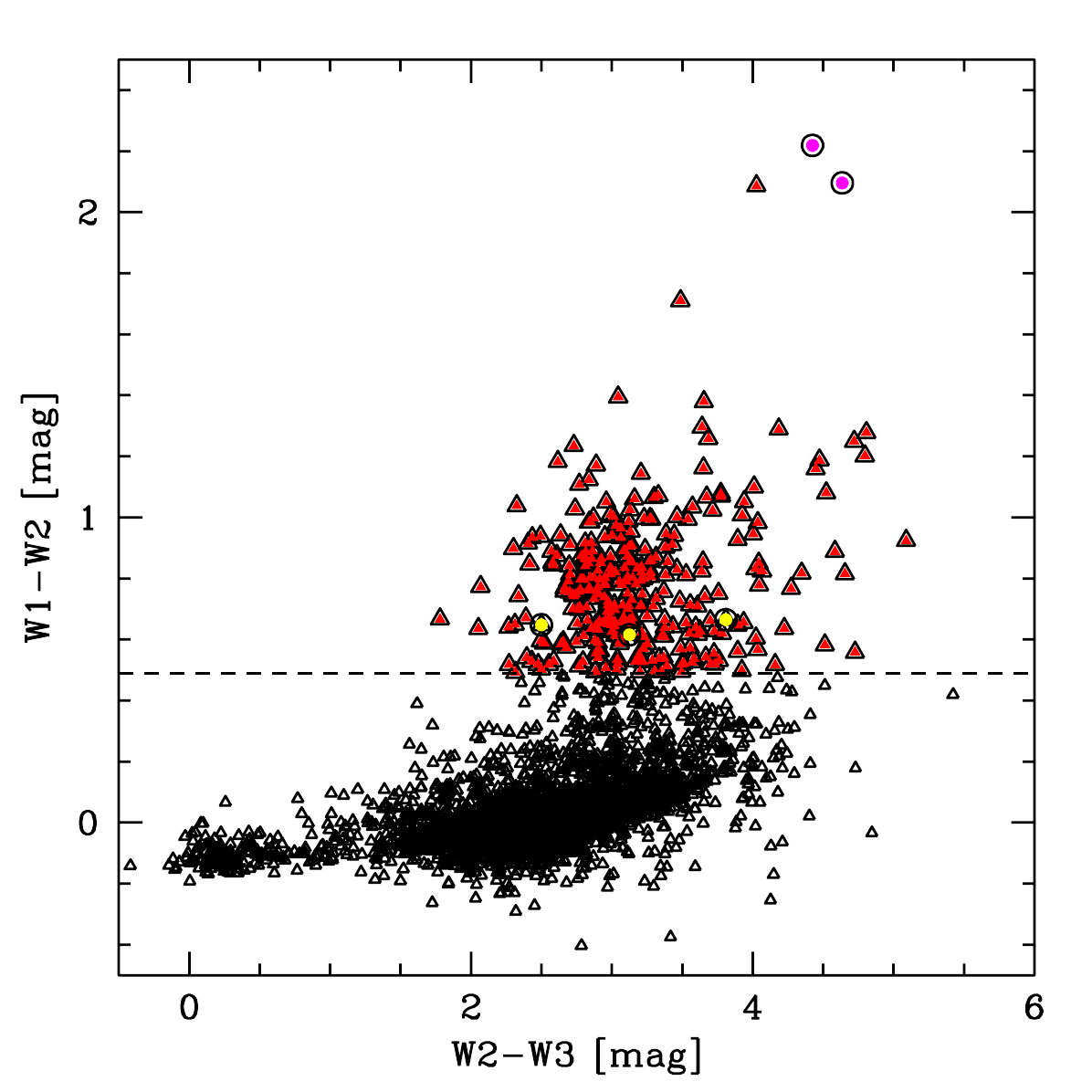}
\includegraphics[width=170pt,height=170pt,angle=0]{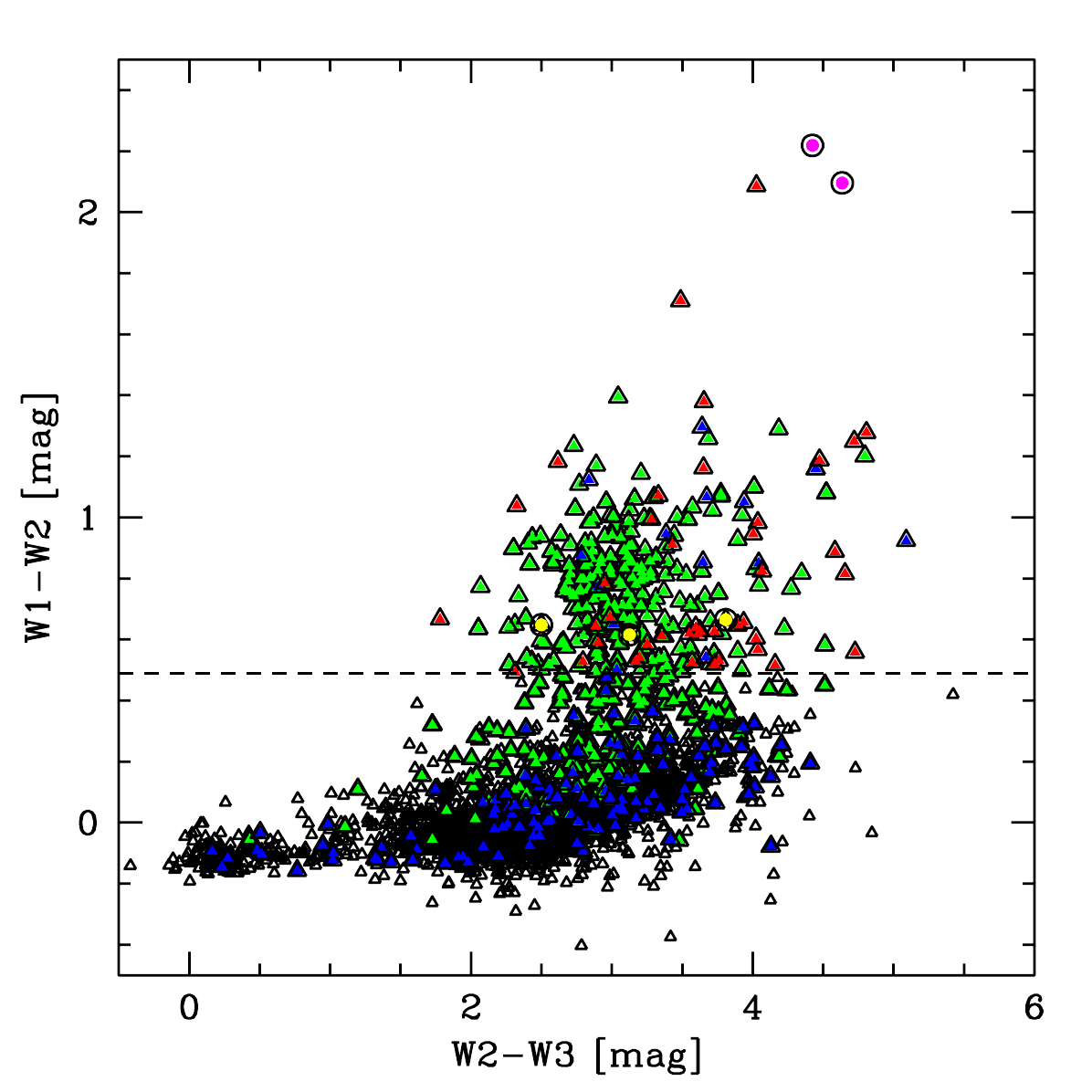}
\includegraphics[width=170pt,height=170pt,angle=0]{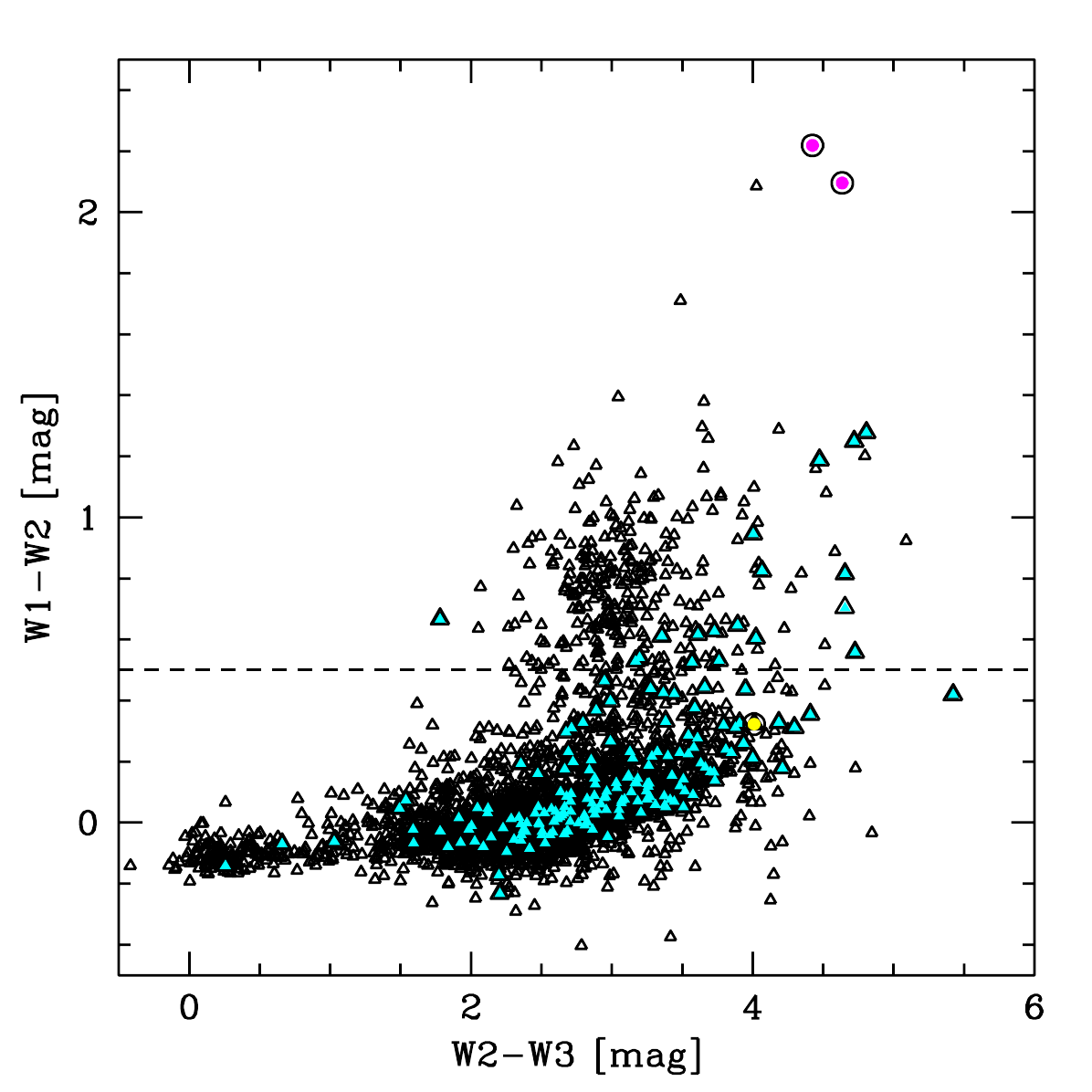}
}
\caption{\label{fig:wisecolor} 
{\it Left:} $W1-W2$ colour vs.\
  $W2-W3$ colour of our 3326 target galaxies ({\it black open
    triangles}). The 303 sources above our IR colour cut are
shown as {\it red filled triangles}. 
The two outliers with
  very red colours in the top right-hand corner of the diagram ({\it
    magenta filled circles}) are the low-metallicity and heavily
  obscured BCDs that were originally identified by \citet{griffith11}.
  The two famous dwarf Seyfert 1 galaxies (NGC~4395 and POX~52), as well
  as a Seyfert 1 galaxy hosting a low-mass BH (UM~625) ({\it yellow
    filled circles}), have WISE colours above our selection
  cut-off. 
{\it Middle:} Same as left, but showing
     the AGN that have been previously identified in the 
  MRBGD optical samples split into 549 type~1 (BL; {\it green
    filled triangles}) and 152 type~2 (NL; {\it blue filled
    triangles}). Although a large fraction of these
  optically identified AGN fall in our IR selected sample of 303
  galaxies, many also have WISE colours below our selection cut-off ({\it
    dashed line}).  The 43 candidate AGN, selected using the IR
  diagnostic, that are not in the samples of MRBGD are shown as {\it
    red filled triangles}. 
{\it Right:} Same as left, but showing the 182 BCDs ({\it cyan
    filled triangles}) and the 2 low-metallicity and heavily
  obscured BCDs that were originally identified by \citet{griffith11}
  ({\it magenta filled circles}). The BCD MRK~709~S ({\it yellow filled
    circle}), one of the most metal-poor BCDs with evidence of an
  active galactic nucleus \citep{reines14}, also has a WISE colour
  below our selection cut-off. \vspace{0.2cm}}
\end{figure*}

\section{Nearby galaxy sample}
\label{sec:sample}

Our primary sources were selected from the Updated Nearby Galaxy
Catalogue of \citet{karachentsev13} and the catalogue of Local Group
(LG) galaxies of \citet{mcconnachie12}. The first is a recently
updated all-sky catalogue containing 869 nearby galaxies with distance
estimates within 11 Mpc or corrected radial velocities less than 600
km s$^{-1}$. The latter contains all known galaxies with distances
determined from measurements of resolved stellar populations that
place them within 3~Mpc of the Sun. We also searched the NASA
Extragalactic Database (NED) under the following three
classifications: dwarfs (dwarf, nucleated, dwarf elliptical (dE),
dwarf lenticular (S0), dwarf spiral (dS), dwarf irregular (dI) and
blue compact dwarf (BCD)); E peculiar; and compact E. 
Dwarfs are ill defined in the literature \citep[see e.g.][for a
review on the various definitions]{dunn10}, and some known dwarfs,
such as NGC~185 and 147, are labelled in NED as elliptical
galaxies. Thus, we expanded our search parameters to ensure that we did not
lose any potential dwarfs in the nearby Universe. Dwarf spirals are a
contentious subject in and of themselves, and we did not search NED
for spiral galaxies to add to the sample. In order to define our dwarf
sample in a consistent manner -- regardless of their previous
classification in the literature -- in the following work, we apply our
own mass cut (see Sections~\ref{sec:IRAGN} and \ref{sec:mstar}).

This primary sample was augmented by the surveys of galaxies,
including dwarfs at the high end of the mass range, that had already
been identified as having the optical spectroscopic signature of
low-mass actively accreting black holes \citep[hereafter
  MRBGD]{moran14,
  reines13,reines14,barth08,greene07,greene04,dong12,dong07}. These
AGN candidates will be used to verify the IR selection technique. The
final list, cleaned of all duplicates, consists of a total of 5897
galaxies (see Table~\ref{tbl:tbl-0} for a summary of the various samples
and their sizes discussed throughout this paper).

\begin{table*}
\centering
\caption{Sample descriptions and sizes. \label{tbl:tbl-0}}
\scriptsize
\begin{tabular}{@{}llr@{}}
\toprule
Name                           &Content                    &Size   \\
\midrule
Primary sample                                   &Catalogue of \citet{karachentsev13};           &5143 \\
						 &catalogue of \citet{mcconnachie12};            & \\
                                                 &NED dwarfs, E peculiar, compact E              & \\
MRBGD samples                         &Catalogues from MRBGD                                     &754 \\
Total sample                              &Primary and MRBGD                                     &5897 \\
\\
WISE sample         	&WISE matches to Primary sample          	 	 		 &5042 \\
WISE S/N sample             &WISE matches to Primary sample with S/N $ > 3$         	 	 &3326 \\
WISE MRBGD AGN sample   &WISE matches to MRBGD NL+BL AGN with S/N $ > 3$                         &700$^a$ \\
WISE MRBGD BL AGN sample   &WISE matches to MRBGD BL AGN with S/N $ > 3$                         &549$^a$ \\
WISE MRBGD NL AGN sample    &WISE matches to MRBGD NL AGN with S/N $ > 3$                        &152$^a$ \\
WISE BCD AGN sample    &WISE matches to BCD AGN with S/N $ > 3$                        		 &182 \\
WISE Griffith AGN sample    &WISE matches to Griffith AGN with S/N $ > 3$                        &2 \\
\\
IR selected AGN sample             &WISE S/N sample with $W1-W2 > 0.5$                               &303 \\
IR new selected AGN sample       &IR selected AGN sample NOT in the MRBGD NL+BL AGN sample       &43 \\
IR selected BCD AGN sample &WISE BCD AGN sample with $W1-W2 > 0.5$                               &18 \\
IR selected MRBGD AGN sample   &WISE MRBGD NL+BL AGN sample with $W1-W2 > 0.5$                   &258$^a$ \\
IR selected MRBGD BL AGN sample   &WISE MRBGD BL AGN sample with $W1-W2 > 0.5$                   &245$^a$ \\
IR selected MRBGD NL AGN sample  &WISE MRBGD NL AGN sample with $W1-W2 > 0.5$                    &15$^a$ \\
IR selected Griffith AGN sample &WISE Griffith AGN sample with $W1-W2 > 0.5$                     &2 \\
IR new selected AGN sample with other AGN diagnostic &IR selected new AGN sample with other AGN diagnostic &16 \\
IR selected AGN sample also identified as AGN via other methods &adding four samples above      &276$^a$ \\
\\
IR selected AGN sample with redshift/distance &IR selected AGN sample with redshift/distance     &300 \\
IR selected new AGN sample with redshift/distance &IR selected new AGN sample with redshift/distance &40 \\
\\
WISE S/N sample in MPH/JHU catalogue    &WISE S/N sample with MPH/JHU catalogue matches                  &968 \\
WISE S/N sample in BPT diagram          &WISE S/N sample with MPH/JHU BPT emission line fluxes           &954 \\
WISE S/N sample with stellar masses    &WISE S/N sample with MPH/JHU stellar masses                      &934 \\
\\
IR selected AGN sample with stellar masses &IR selected AGN sample with MPH/JHU stellar masses   &264 \\
IR selected dwarf AGN sample              &IR selected AGN sample with log M$_{stellar} < 9.5$   &62 \\  
IR selected AGN sample with stellar masses and redshift/distance &IR selected AGN sample with MPH/JHU stellar masses and redshift/distance                &264 \\
IR selected MRGBD AGN sample with stellar masses and redshift/distance &IR selected BL+NL AGN sample with MPH/JHU stellar masses and redshift/distance    &239$^a$ \\
IR selected MRGBD BL AGN sample with stellar masses and redshift/distance &IR selected BL AGN sample with MPH/JHU stellar masses and redshift/distance    &227$^a$ \\
IR selected MRGBD NL AGN sample with stellar masses and redshift/distance &IR selected NL AGN sample with MPH/JHU stellar masses and redshift/distance    &14$^a$ \\
IR selected Griffith AGN sample with stellar masses and redshift/distance &IR selected Griffith AGN sample with MPH/JHU stellar masses and redshift/distance  &2 \\
IR selected new AGN sample with stellar masses and redshift/distance &IR selected new AGN sample with MPH/JHU stellar masses and redshift/distance            &23 \\
IR selected NED BCD AGN sample with stellar masses and redshift/distance &IR selected NED BCD AGN sample with MPH/JHU stellar masses and redshift/distance    &15 \\
\\
IR selected AGN sample in BPT diagram &IR selected AGN sample with MPH/JHU BPT emission line fluxes                       &135 \\
IR selected MRBGD AGN sample in BPT diagram   &IR selected MRBGD NL+BL AGN sample with MPH/JHU BPT emission line fluxes   &126$^a$ \\
IR selected MRBGD NL AGN sample in BPT diagram  &IR selected MRBGD BL AGN sample with MPH/JHU BPT emission line fluxes    &111$^a$ \\
IR selected MRBGD BL AGN sample in BPT diagram  &IR selected MRBGD NL AGN sample with MPH/JHU BPT emission line fluxes    &16$^a$ \\
IR selected new AGN sample in BPT diagram &IR selected new AGN sample with MPH/JHU BPT emission line fluxes               &9 \\
IR selected NED BCD AGN sample in BPT diagram &IR selected NED BCD AGN sample with MPH/JHU BPT emission line fluxes       &83 \\
\\
IR selected AGN sample with IR and [OIII]~$\lambda$5007 BH masses &IR selected AGN sample with IR BH masses and [OIII]~$\lambda$5007 BH masses            &248 \\
\bottomrule
\end{tabular}
\tablefoot{
$^a$ Two objects in this sample are classified as both NL and BL.
}
\end{table*}

\begin{figure*}
\centerline{
\includegraphics[width=500pt,height=559pt,angle=0]{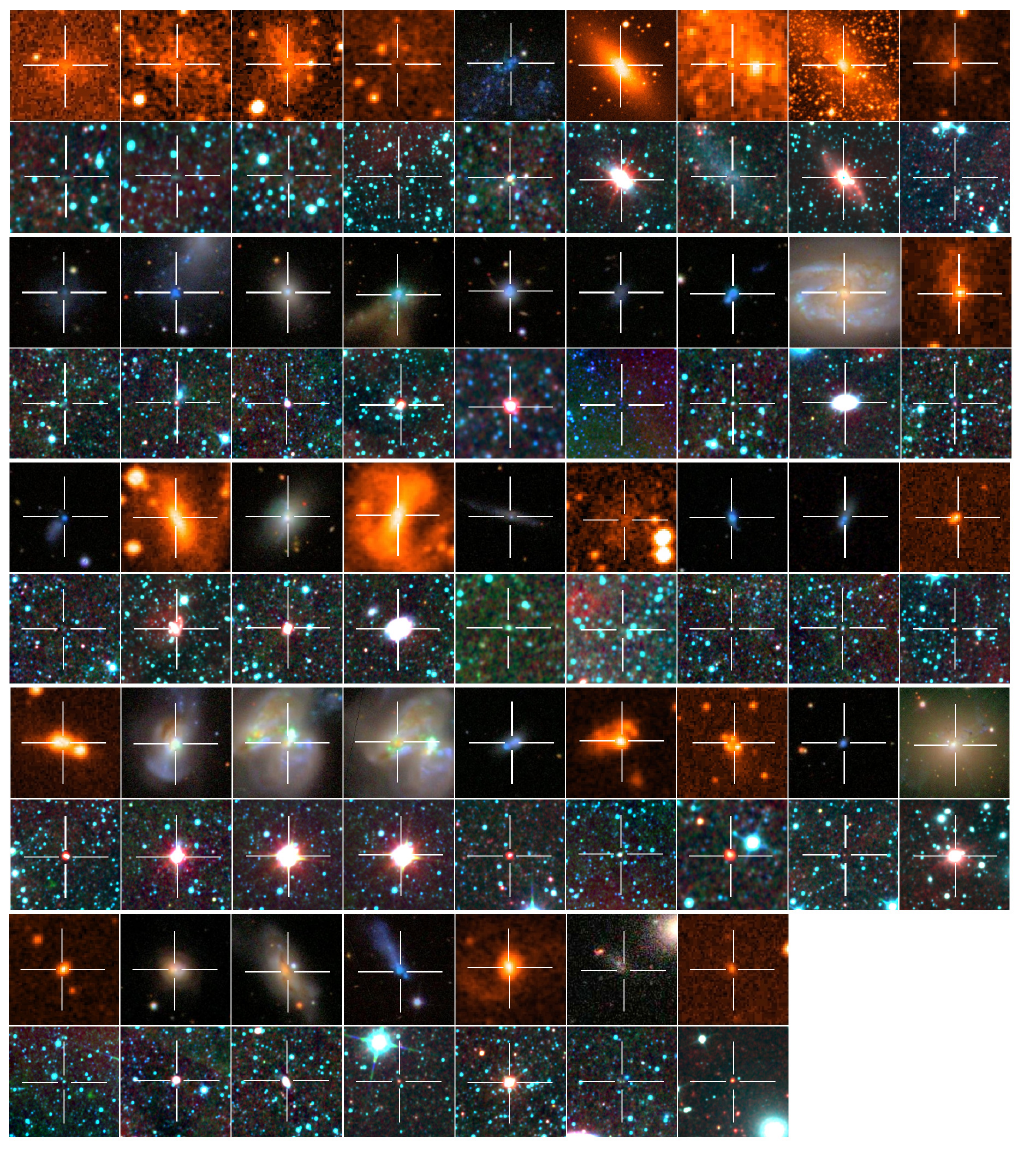}
}
\caption{\label{fig:postagestamps} Mosaic of the 43
    candidate AGN, selected using the IR diagnostic, which are not in
    the samples of MRBGD. They are ordered, from {\it left} to {\it right}
  and {\it top} to {\it bottom}, by increasing distance with the
  exception of the first three candidates, which have no known
  redshift/distance (see Table~\ref{tbl:tbl-1}). The {\it top} images
  are either optical g',r',i' colour images from SDSS, which are 76.8''
  on a side (except for OBJ 36, which is 153'' on a side), or
  false colour images from the DSS-2-red catalogue, which are 1' on a
  side (except for OBJ 6 and 8, which are 4' on a side). The {\it
    bottom} images are W1,W2,W3 colour images from WISE, $\sim$ 9' on a
  side (except for OBJ 14, 23 and 34, which are $\sim$ 4' on a
  side). \vspace{0.2cm}}
\end{figure*}

\begin{figure*}
\centerline{
\includegraphics[width=170pt,height=170pt,angle=0]{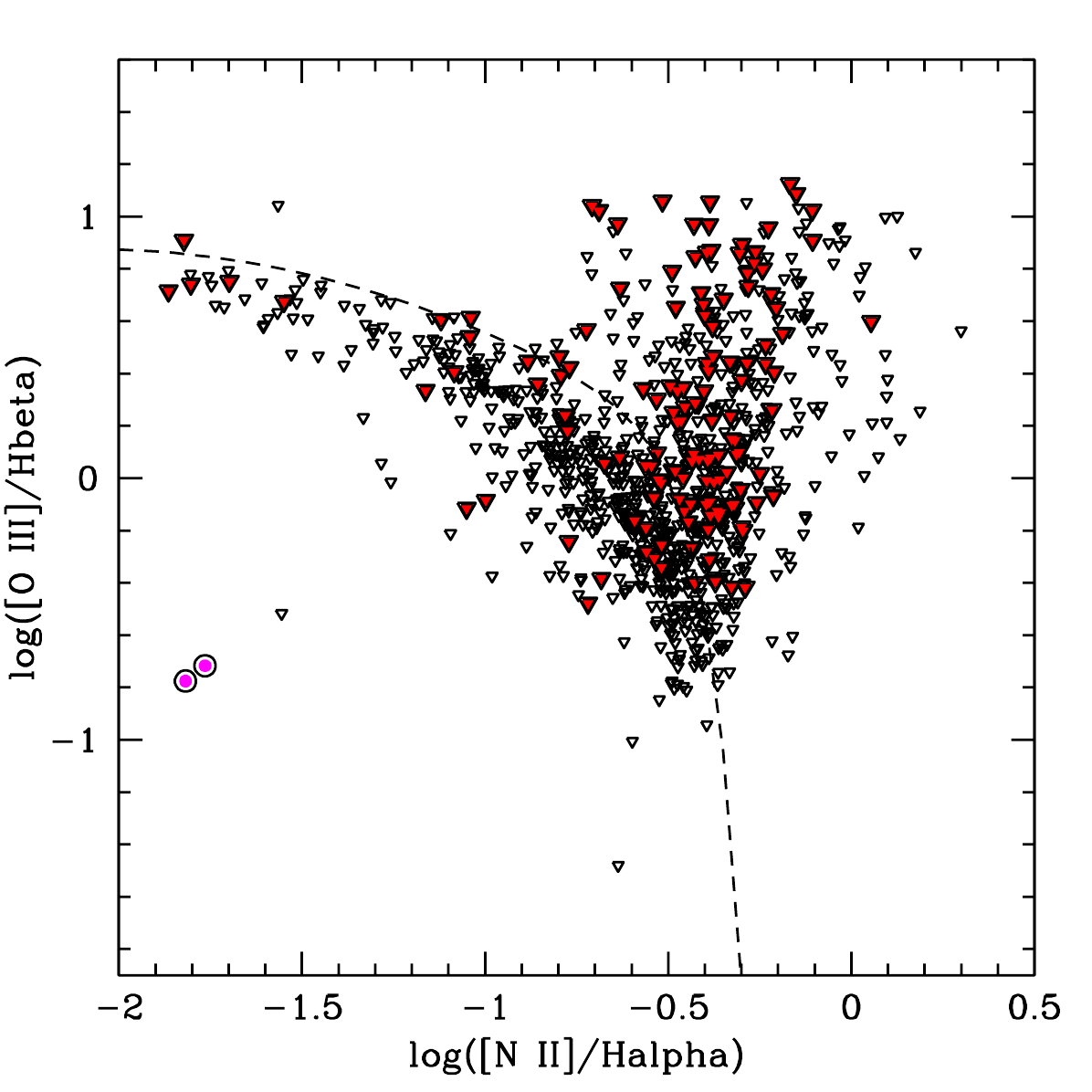}
\includegraphics[width=170pt,height=170pt,angle=0]{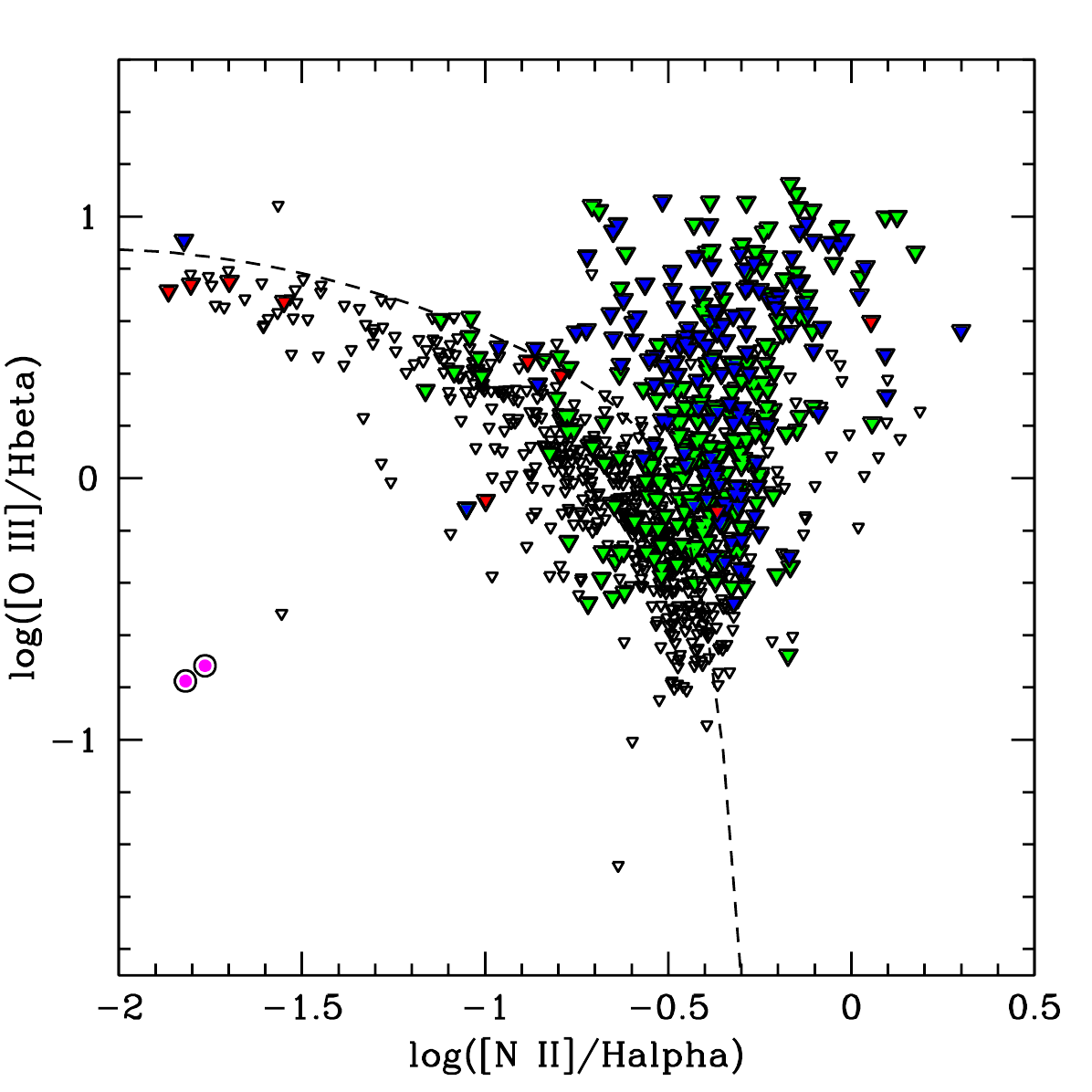}
\includegraphics[width=170pt,height=170pt,angle=0]{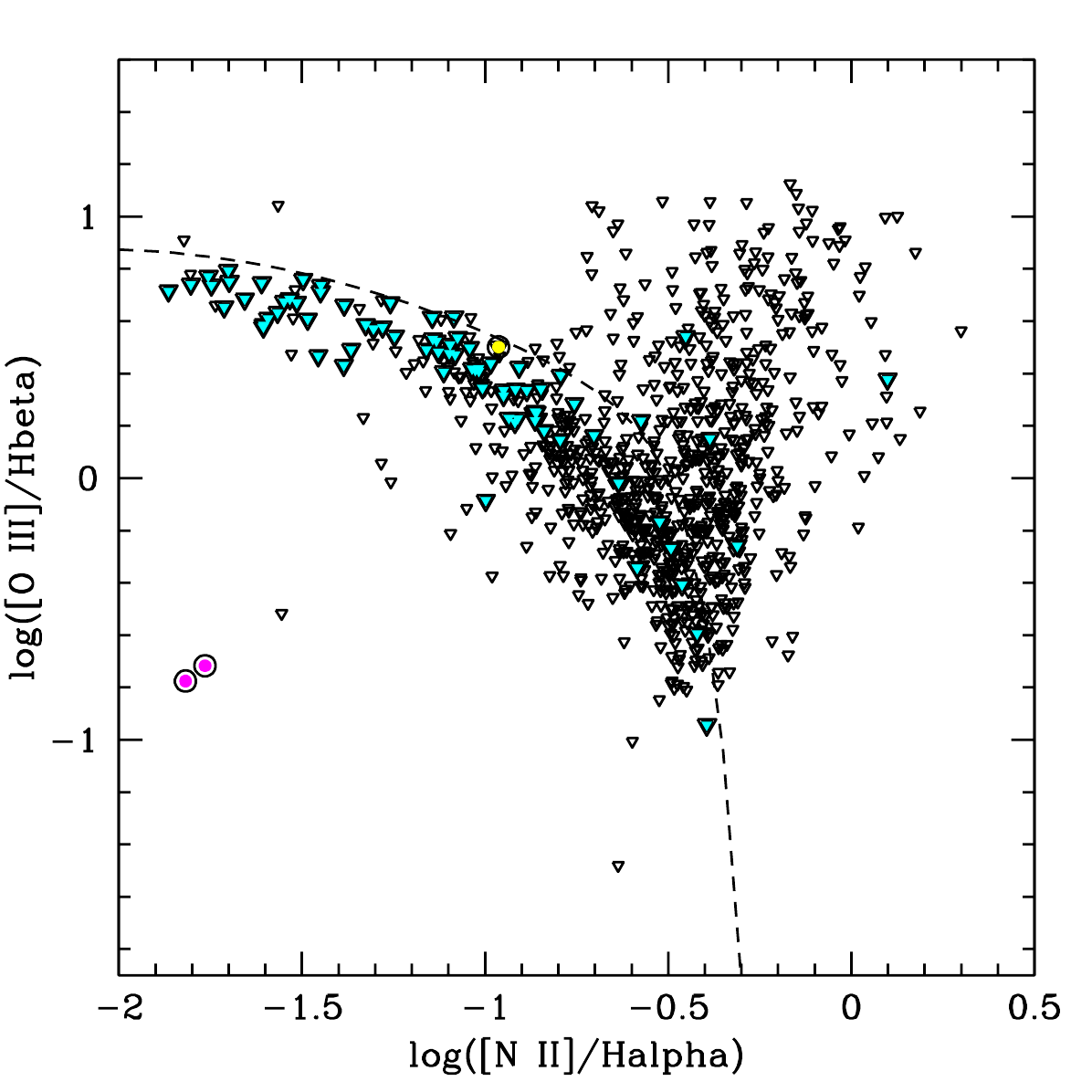}
}
\caption{\label{fig:bpt} 
{\it Left:} [OIII]/H$\beta$ vs.\
  [NII]/H$\alpha$ diagnostic diagram for the 954 out of the 3326
  galaxies with both line flux ratio measurements from the MPA/JHU
  catalogue. As in Figure~\ref{fig:wisecolor}, the 135 sources above our
IR colour cut are shown as {\it red filled triangles}.
The {\em dashed line} is the
  demarcation line between normal star-forming galaxies and AGN from
  \citet{stasinska06}.
{\it Middle:} Same as left, but showing
     the 126 AGN that were previously identified in the 
   MRBGD optical samples split into 111 type~1 (BL; {\it green
    filled triangles}) and 16 type~2 (NL; {\it blue filled
    triangles}). One-third of this MRBGD sample are located
  in the region for HII galaxies. The 9 candidate AGN, selected using the IR
  diagnostic, which are not in the samples of MRBGD are shown as {\it
    red filled triangles}.
{\it Right:} Same as left, but showing the 83 BCDs ({\it cyan
    filled triangles}) and the two low-metallicity and heavily
  obscured BCDs that were originally identified by \citet{griffith11}
  ({\it magenta filled circles}). The BCD MRK~709~S ({\it yellow filled
    circle}), one of the most metal-poor BCDs with X-ray and radio emission
indicative of AGN activity, is also located in the region for HII galaxies. \vspace{0.2cm}}
\end{figure*}

\section{Mid-infrared colours of nearby galaxies}
\label{sec:IRcolor}

For this work, we used mid-infrared colours to identify the AGN in our
sample. One advantage of using images in the mid-infrared is that 
both unobscured (type~1) and obscured (type~2) AGN are detected. 
The mid-infrared selection of unobscured AGN relies upon
distinguishing the approximately power-law AGN spectrum from the 
black-body stellar spectrum of galaxies \citep[which peaks at rest-frame 
1.6 $\mu$m;][Figure 3]{assef10} using its red mid-infrared colours. It is
important to note that most of the previous surveys of AGN in dwarf galaxies 
have been carried out in the
optical. Optical studies are biased towards observing primarily type~1
AGN, even though the unified AGN model \citep{antonucci93,urry95}
predicts that type~2 AGN should outnumber type~1 by a factor of
$\sim$3 \citep[e.g.,][]{comastri95, treister04,
  ballantyne11}. Another advantage to working with mid-infrared data
is that they allow AGN to be easily distinguished from stars and
galaxies.

The infrared magnitudes of our galaxies were obtained from the
Wide-field Infrared Survey Explorer \citep[WISE;][]{wright10} All-Sky
Release Source Catalog. This catalogue contains positions and photometry
at 3.4 (W1), 4.6 (W2), 12 (W3) and 22 (W4) $\mu$m for 563,921,584
point-like and resolved objects detected on the Atlas Intensity
images. The photometry in this catalogue was performed using point
source profile-fitting and multi-aperture photometry and the estimated
sensitivities are 0.068, 0.098, 0.86 and 5.4 mJy (5$\sigma$) at 3.4,
4.6, 12 and 22 $\mu$m in unconfused regions on the ecliptic
plane. J2000 positions and uncertainties were reconstructed using the
2MASS Point Source Catalog as astrometric reference. Astrometric
accuracy is approximately 0.2 arcsec root-mean-square on each axis
with respect to the 2MASS reference frame for sources with
signal-to-noise ratio (S/N) greater than forty.

We cross-matched our nearby galaxy catalogue with the WISE All-Sky
Release Source Catalog using the US Virtual Astronomical Observatory
(VAO) cross-comparison tool. Using a matching radius of 6 arcseconds,
corresponding to the resolution of WISE in W1, we obtained 5042
matches. For each of the galaxies, WISE photometry for W1, W2, W3 and
W4 was gathered and a S/N cut of 3 was imposed on
the first three bands since only these three bands are needed for the
AGN diagnostic diagram and to obtain an estimate of the BH mass (see
Section~\ref{sec:bhmass}). We used the point source profile-fitting
photometry (w1mpro, w2mpro, w3mpro and w4mpro) for sources with
goodness-of-fit $\leq$ 3.0 which indicates that the source shape is
consistent with a point source and the source is not associated with
or superimposed on a 2MASS Extended Source Catalog (XSC) source. For
sources with goodness-of-fit $>$ 3.0, we used the 5.5 arcsec radius
aperture photometry (w1mag\_1, w2mag\_1, w3mag\_1 and w4mag\_1) to
measure the IR colours. Using the three photometric bands, two
photometric colours were determined. In Figure~\ref{fig:wisecolor}, we
plot the $W1-W2$ versus $W2-W3$ colour for the 3326 galaxies in our
final S/N $ > 3$ WISE matched sample.

\section{Infrared colour diagnostic for AGN candidates}
\label{sec:IRAGN}

Several mid-infrared colour diagnostics have been used in the
literature to select AGN, starting with the pioneering work of
\citep{lacy04,lacy07} and the so-called ``Lacy wedge''. For low
redshift galaxies (i.e.\ $z < 1.3$), which applies to our sample of
nearby galaxies, \citet{stern12} show that WISE is able to robustly
identify AGN using the more sensitive W1 (3.4 $\mu$m) and W2 (4.6
$\mu$m) bands alone. According to these authors, a colour cut of $W1-W2
> 0.8$, 0.7, 0.6, and 0.5 is able to identify AGN with a reliability
of 95\%, 85\%, 70\% and 50\%, respectively. Note that although
reliability drops with lower colour cut, completeness increases
and reaches the 95\% level for $W1-W2 > 0.5$. As can be seen in their
Figure 2, even the least stringent colour cut of $W1-W2 > 0.5$
identifies galaxies in which the AGN fraction is $\geq 50$\% to $z =
0.5$ (this fraction changes depending on extinction). This
means that this colour cut only identifies AGN that dominate the
emission from their host galaxies. For galaxies in which the AGN do
not dominate over the host galaxy emission, dilution of the
mid-infrared AGN continuum by the host galaxy light can cause bluer
$W1-W2$ colours that become indistinguishable from star-forming
galaxies. In fact, as is shown in Section~\ref{subsec:bptdiag}, a
large fraction of the optically identified AGN in the Sloan Digital
Sky Survey (SDSS) galaxy samples of MRBGD have $W1-W2$ colours below
this cut-off.

The 3326 galaxies with WISE matches and S/N $ > 3$ in $W1$, $W2$ and
$W3$ are shown in Figure~\ref{fig:wisecolor}, left. With our colour cut
of $W1-W2 > 0.5$, we identify 303 candidate AGN (see
Figure~\ref{fig:postagestamps}).  These final numbers come after we
conservatively removed candidates that were found close to an image
artefact that could possibly contaminate the source and give an
erroneous photometry. Of the 303 candidate AGN, 62 are classified as
dwarf galaxies based on a $\log M_{stellar}$ cut of 9.5 \citep[similar
  to ][]{reines13}. The list of these galaxies, ordered by increasing
distance, can be found in Tables~\ref{tbl:tbl-1} (dwarfs) and
\ref{tbl:tbl-2} (non-dwarfs).

\begin{table*}
\centering
\caption{List of dwarf galaxies with IR signatures of active massive black holes. \label{tbl:tbl-1}}
\small
\begin{tabular}{@{}llrrrrrrrr@{}}
\toprule
Name                           &RA           &Dec          &Redshift &Distance  &log(M$_{stellar}$)     &W1-W2    &W2-W3    &log(M$_{BH\;IR}$) & Image \\
                               &J2000        &J2000        &         &Mpc       &M$_{\odot}$                &[mag]    &[mag]    &M$_{\odot}$      &       \\
\midrule
                   AM 1906-621  &19h11m33.43s  & -62d10m22.4s &     -        &    -     &    -     &    0.784 &   2.950  &   -     &  1 \\
                   AM 1238-405  &12h41m15.40s  & -41d09m34.0s &     -        &    -     &    -     &    0.628 &   3.626  &   -     &  2 \\
                ESO 184- G 050  &19h16m17.23s  & -54d20m40.8s &     -        &    -     &    -     &    0.586 &   3.252  &   -     &  3 \\
               HIPASS J1247-77  &12h47m32.60s  & -77d35m01.0s &     0.001378 &    3.160 &    -     &    2.086 &   4.026  &   2.192 &  4 \\
                     UGC 04459  &08h34m07.20s  & +66d10m54.0s &     0.000067 &    3.253 &    -     &    1.379 &   3.654  &   2.687 &  5 \\
                       NGC5253  &13h39m55.80s  & -31d38m24.0s &     0.001358 &    3.646 &    -     &    1.710 &   3.487  &   5.602$^a$ &  6 \\
                       IC 2574  &10h28m23.48s  & +68d24m43.7s &     0.000190 &    3.835 &    -     &    0.646 &   2.885  &   1.638 &  7 \\
                      NGC 4395  &12h25m48.86s  & +33d32m48.7s &     0.001064 &    4.491 &    7.156 &    0.648 &   2.500  &   3.589 &  - \\
               HIPASS J1337-39  &13h37m25.10s  & -39d53m52.0s &     0.001641 &    4.800 &    -     &    0.888 &   4.583  &   2.506 &  9 \\
                     UGC 07544  &12h26m37.74s  & +62d22m47.2s &     0.002372 &    9.499 &    6.728 &    0.590 &   2.902  &   2.569 & 10 \\
                      MRK 0094  &08h37m43.48s  & +51d38m30.3s &     0.002451 &    9.817 &    5.980 &    0.816 &   4.655  &   3.839 & 11 \\
                       UGCA298  &12h46m55.40s  & +26d33m51.0s &     0.002743 &    9.950 &    8.175 &    0.519 &   3.730  &   4.758 & 12 \\
                      UGCA 116  &05h55m42.60s  & +03d23m32.0s &     0.002632 &   10.300 &    7.286 &    1.278 &   4.806  &   5.746 & 13 \\
\bottomrule
\end{tabular}
\tablefoot{
Col. (1): Object Name. Cols. (2) and (3);
Right ascension and declination (J2000). Col. (4): Redshift. Col. (5):
Distance in Mpc. Col. (6): Log of host galaxy stellar mass measured from SED
fit. Cols. (7) and (8): WISE colours. Col. (9): Log of central black hole mass 
estimates based on IR luminosity (assuming $L_{bol} = 0.1 L_{Edd}$). 
Col. (10): Image number in Figure~\ref{fig:postagestamps}.
\vskip 4pt
$^a$ These sources are extended and larger than the largest WISE aperture (24.75 arcsec radius aperture); 
therefore, the infrared black hole mass estimates for these sources are only lower limits.
\vskip 4pt
Table~\ref{tbl:tbl-1} is published in its entirety in the
electronic edition of the A\&A. A portion is shown here for guidance
regarding its form and content.

}
\end{table*}

In our analysis, we use the colour cut of $W1-W2 > 0.5$, without
imposing any $W2-W3$ colour restriction. One cause for possible concern
when using this cut is the contamination of low metallicity star-forming
galaxies. The effect of metallicity on the intensity of the
polycyclic aromatic hydrocarbon (PAH) emission has been quantified in
detail using Spitzer data \citep[e.g.][and references
  therein]{calzetti11}.  Analyses of galaxies with a range of
metallicities show that a factor of $\sim$10 decrease in metallicity is
accompanied by an order of magnitude decrease in the 8 $\mu$m to total
infrared luminosity (i.e.\ redder $W2-W3$ colours), with a transition
at 12+log(O/H)$\approx$8.1
\citep{boselli04,madden06,engelbracht05,hogg05,galliano05,galliano08,rosenberg06,wu06,draine07,engelbracht08,gordon08,munoz09,marble10}. Based
on the galaxy mass-metallicity relation \citep{ma16}, this metallicity
transition corresponds to a stellar mass of $\log M_{stellar} \sim
8.0$. Hence many dwarfs, based on our mass cut-off ($\log M_{stellar}
= 9.5$), are not expected to have such low metallicity
or redder $W2-W3$ colours. The zero-point of the mass-metallicity
relation changes with redshift, so the colours of low-metallicity
dwarfs becomes more of an issue as one moves to higher redshifts.

In addition to the confusion between star-forming low-metallicity
galaxies and AGN, there is growing evidence that nuclear star clusters
can coincide with central massive black holes
\citep[e.g.][]{denBrok15,georgiev16}.  In the case of NGC~4395, the
$4 \times 10^5$ M$_{\odot}$ central black hole is embedded in a
nuclear star cluster of mass $2 \times 10^6$ M$_{\odot}$
\citep{denBrok15}, and is detected via its infrared signature
\citep{satyapal14}. Hence, strong levels of star formation do not
negate the possible presence and IR detection of an AGN.

\begin{table*}
\centering
\caption{List of non-dwarf galaxies with IR signatures of active massive black holes. \label{tbl:tbl-2}}
\small
\begin{tabular}{@{}llrrrrrrrr@{}}
\toprule
Name                           &RA           &Dec          &Redshift &Distance  &log(M$_{stellar}$)     &W1-W2    &W2-W3    &log(M$_{BH\;IR}$) & Image \\
                               &J2000        &J2000        &         &Mpc       &M$_{\odot}$                 &[mag]    &[mag]    &M$_{\odot}$         &       \\
\midrule
                      CIRCINUS  &14h13m09.30s  &-65d20m21.0s  &    0.001448  &   4.207  &   -      &   1.038  &   2.325  &   6.306$^a$ &  8 \\
                      NGC 2964  &09h42m54.23s  &+31d50m50.6s  &    0.004430  &  20.429  &  10.201  &   0.667  &   1.780  &   6.497$^a$ & 17 \\
                      NGC 4194  &12h14m09.47s  &+54d31m36.6s  &    0.008178  &  39.100  &   9.684  &   0.531  &   3.761  &   7.235$^a$ & 29 \\
                NGC 3690 NED01  &11h28m31.02s  &+58d33m40.7s  &    0.009990  &  40.240  &  10.357  &   1.183  &   2.617  &   7.877$^a$ & 30 \\
                NGC 3690 NED02  &11h28m33.63s  &+58d33m46.6s  &    0.010456  &  42.133  &  10.461  &   0.983  &   4.036  &   7.864$^a$ & 31 \\
                       Moran08  &10h05m51.19s  &+12d57m40.6s  &    0.009375  &  43.300  &   9.777  &   1.124  &   2.838  &   6.466 &  - \\
                      NGC 1275  &03h19m48.16s  &+41d30m42.1s  &    0.017559  &  68.180  &  11.067  &   0.911  &   3.429  &   7.752$^a$ & 36 \\
             Greene137+Dong164  &11h53m41.77s  &+46d12m42.2s  &    0.024251  &  98.744  &  10.329  &   0.518  &   2.270  &   6.576 &  - \\
                       Dong222  &14h00m40.57s  &-01d55m18.3s  &    0.025049  & 102.052  &   9.661  &   0.665  &   3.809  &   6.678 &  - \\
                      Greene47  &08h24m43.28s  &+29d59m23.5s  &    0.025420  & 103.594  &   9.674  &   1.171  &   2.888  &   7.379 &  - \\
                     Reines3NL  &03h22m24.64s  &+40d11m19.8s  &    0.026084  & 106.352  &   9.565  &   0.854  &   3.647  &   5.988 &  - \\
                      MRK 0315  &23h04m02.62s  &+22d37m27.5s  &    0.027278  & 111.319  &  10.854  &   0.527  &   2.796  &   7.259 & 38 \\
                     Greene203  &15h59m09.62s  &+35d01m47.4s  &    0.031024  & 126.958  &  10.600  &   0.743  &   2.338  &   7.279$^a$ &  - \\
\bottomrule
\end{tabular}
\tablefoot{
Col. (1): Name. Cols. (2) and (3);
Right ascension and declination (J2000). Col. (4): Redshift. Col. (5):
Distance in Mpc. Col. (6): Log of stellar mass measured from SED
fit. Cols. (7) and (8): WISE colours. Col. (9): Log of central black hole mass 
estimates based on IR luminosity (assuming $L_{bol} = 0.1 L_{Edd}$). 
Col. (10): Image number in Figure~\ref{fig:postagestamps}.
\vskip 4pt
$^a$ These sources are extended and larger than the largest WISE aperture (24.75 arcsec radius aperture); 
therefore, the infrared black hole mass estimates for these sources are only lower limits.
\vskip 4pt
Table~\ref{tbl:tbl-2} is published in its entirety in the
electronic edition of the A\&A. A portion is shown here for guidance
regarding its form and content. 
}
\end{table*}

If we were to apply a cut on the $W2-W3$ colour to avoid the redder
objects, e.g.\ a cut with $W2-W3 < 4.2$ (similar to equation 1 of
\citealt{jarrett11}), in addition to our $W1-W2$ cut, we would reject 16
AGN candidates (see Figure~\ref{fig:wisecolor}). Of these, 2 are the
low-metallicity and heavily obscured BCDs that were originally
identified by \citet{griffith11}, but 8 are either type 1 (2) or type 2 (6)
optically identified AGN. We note that the remaining 6 that were not
previously identified as AGN have colours similar to these 8
optically identified AGN and hence are valid candidate AGN. We also note
that AGN detected in dwarf galaxies via other methods, such as the
dwarf galaxy Henize 2-10 \citep{reines11}, have $W2-W3$ colours $> 4.2$
($W2-W3 \sim 5.0$ for Henize 2-10; \citealt{satyapal14}) and do not fall
within the \citet{jarrett11} demarcation.

\begin{figure}
\centerline{
\includegraphics[width=240pt,height=240pt,angle=0]{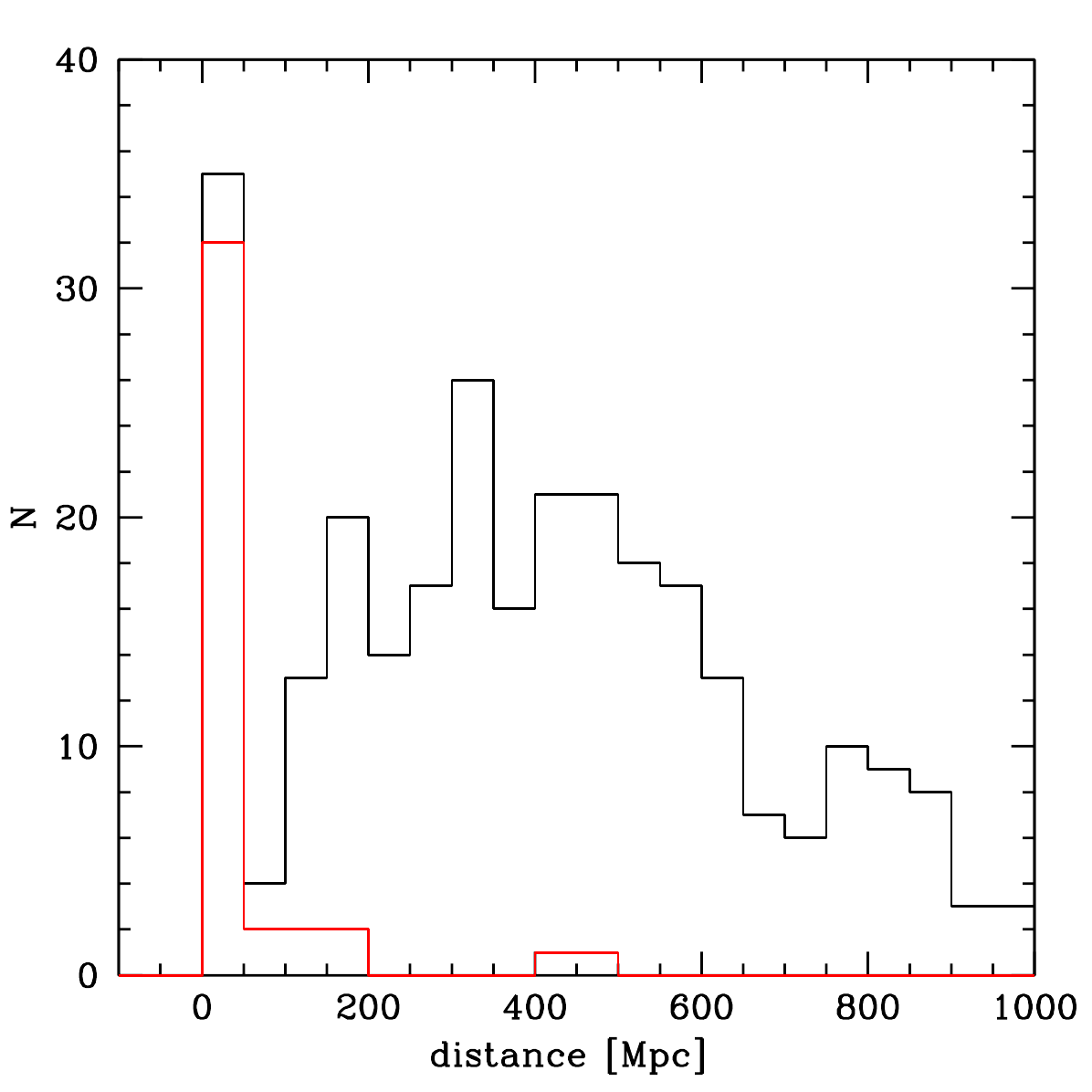}
}
\caption{\label{fig:hist_dist} Distance distribution of our IR selected sample of
  300 AGN candidates with known redshift/distance. The subsample 
of 40 galaxies selected via the IR diagnostic, shown here in red, contains 
the nearest AGN candidates known today. \vspace{0.2cm}}
\end{figure}

\section{Verification using existing catalogues and optical line emission diagnostic}
\label{sec:verification}

\subsection{Existing catalogues}
\label{subsec:cat}

To verify the validity of our method, we compare our final list
of AGN candidates with AGN candidates determined by other methods.
The results of this comparison are listed below.

\begin{table*}
\centering
\caption{List of 16 known AGN with IR signatures of active massive black holes. \label{tbl:tbl-3}}
\small
\begin{tabular}{@{}llrrrrr@{}}
\toprule
Name                           &RA           &Dec          &Classification &log(M$_{BH\;IR}$) &log(M$_{BH}$) & Reference \\
                               &J2000        &J2000        &               &M$_{\odot}$         &M$_{\odot}$     &       \\
\midrule
                      NGC 5253  &13h39m55.80s  &-31d38m24.0s  &Sy2       &5.602$^a$     &-           &Koulouridis (2014) \\
                      CIRCINUS  &14h13m09.30s  &-65d20m21.0s  &Sy2       &6.306$^a$     &6.23        &McConnell et al.\ (2013) \\
                      NGC 2964  &09h42m54.23s  &+31d50m50.6s  &-         &6.497$^a$     &7.34;6.11   &Beifiori et al.\ (2012) \\
                ESO 060- G 019  &08h57m26.72s  &-69d03m36.3s  &-         &6.015$^a$     &7.83        &Davis et al.\ (2014) \\
                       IC 1953  &03h33m41.87s  &-21d28m43.1s  &-         &5.990$^a$     &7.33        &Davis et al.\ (2014) \\
                      NGC 4194  &12h14m09.47s  &+54d31m36.6s  &AGN       &7.235$^a$     &-           &SIMBAD info page \\
                NGC 3690 NED01  &11h28m31.02s  &+58d33m40.7s  &Sy2       &7.877$^a$     &7.52        &Alonso-Herrero et al.\ (2014) \\
                NGC 3690 NED02  &11h28m33.63s  &+58d33m46.6s  &LINER     &7.864$^a$     &7.48;8.85   &Alonso-Herrero et al.\ (2014); \\
                                &              &              &          &          &            &Zhao et al.\ (1997) \\
               IRAS 11485-2018  &11h51m11.60s  &-20d36m02.0s  &AGN       &5.591     &-           &Sargsyan et al.\ (2011) \\
                      NGC 1275  &03h19m48.16s  &+41d30m42.1s  &LINER;Sy1 &7.752$^a$     &9.28;8.51   &McKernan et al.\ (2010); \\
                                &              &              &          &          &            &Panessa et al.\ (2006) \\
                        POX052  &12h02m56.91s  &-20d56m02.7s  &Sy1       &5.597     &5.2         &Barth et al.\ (2004) \\
                      MRK 0315  &23h04m02.62s  &+22d37m27.5s  &Sy1       &7.259     &-           &Veron-Cetty \& Vernon (2010) \\
                     UGC 04211  &08h04m46.38s  &+10d46m36.2s  &Sy2       &7.156     &-           &Veron-Cetty \& Vernon (2010) \\
                  SBS 1415+437  &14h17m01.41s  &+43d30m05.5s  &QUASAR    &5.934     &-           &Bukhmastova (2001) \\
                ESO 253- G 003  &05h25m18.08s  &-46d00m21.0s  &Sy2       &8.090$^a$     &-           &Veron-Cetty \& Vernon (2010) \\
         FCSS J033846.0-352252  &03h38m45.97s  &-35d22m52.3s  &Sy2       &7.281     &-           &Veron-Cetty \& Vernon (2010) \\
\bottomrule
\end{tabular}
\tablefoot{In Col. (5), $M_{BH\;IR}$ denotes the central black hole mass estimates
based on IR luminosity (assuming $L_{bol} = 0.1 L_{Edd}$), 
while in Col. (6), $M_{BH}$ refers to the central black hole mass estimates 
based on other conventional (non-infrared) methods.
\vskip 4pt
$^a$ These sources are extended and larger than the largest WISE aperture (24.75 arcsec radius aperture); 
therefore, the infrared black hole mass estimates for these sources are only lower limits.
}
\end{table*}

\begin{enumerate}

\item We find that 258 of the previously optically identified AGN in
  the MRBGD samples are also above our IR colour cut (see
  Figure~\ref{fig:wisecolor}).  Of the 43 that are not in the MRBGD
  samples, a literature search revealed that 16 had previously been
  identified as hosting an AGN (see Table~\ref{tbl:tbl-3}), further
  supporting our IR colour selection criterion for a total of 276 out
  of 303 (or 91\%) AGN candidates.

\item By splitting the MRBGD samples into type~1 (broad-line, hereafter
  BL) and type~2 (narrow-line, hereafter NL), we can see that type~1
  AGN appear on average to have redder $W1-W2$ IR colours and lie
  preferentially above the IR cut-off, as compared to the type~2 AGN
  which lie preferentially below. This simply demonstrates what was
  discussed in Section~\ref{sec:IRcolor} above, that the IR colour
  diagnostic does not exhaustively pick out AGN as their signatures
  can be washed out by an appreciable level of star-forming activity.

\item In Figure~\ref{fig:wisecolor}, we highlight three famous Seyfert
  1 galaxies, NGC~4395, POX~52, and UM~625, to show where they fall
  on the diagram ({\it yellow circles}). As discussed above, NGC~4395
  has the only dynamical mass measurement of a CMBH in a dwarf galaxy
  \citep{denBrok15}, POX~52 has a robust CMBH mass estimate
  \citep{barth04} and UM~625 is a Seyfert 1 galaxy hosting a low-mass
  CMBH \citep{jiang13}. It should be noted that UM~625 was
  added manually, and is not part of our sample as it is neither a
  dwarf nor nearby, though its low-mass CMBH still makes it an object
  of interest. We see that the AGN candidates for these three galaxies 
  have colours above our selection cut-off.

\item It can be noted that there are four outliers with very red
  colours, i.e.\ $W1-W2 > 1.7$, in the top right-hand corner of the
  diagram. Two of these ({\it magenta circles} in
  Figure~\ref{fig:wisecolor}) are the low-metallicity and heavily
  obscured BCDs that were originally identified by \citet{griffith11}
  as already discussed above. These have been flagged in
  Table~\ref{tbl:tbl-1} as they are outliers and have optical
  diagnostics consistent with HII regions. The other two are
  HIPASS~J1247-77 ($W1-W2 \sim 2.086$) and NGC~5253 ($W1-W2 \sim
  1.757$), one of the nearest known BCDs. For these, we have no optical
  diagnostic information.

\item In Figure~\ref{fig:wisecolor}, right, we show where the NED
  sample of BCD galaxies are located. BCDs are small, gas rich
  galaxies that are currently in a period of enhanced star
  formation. These galaxies are characterized by their blue colour, low
  metallicity and compact size.  Some are extremely dusty
  \citep{griffith11} ({\it magenta circles} found in the top left
  corner of Figure~\ref{fig:wisecolor}, right) and appear to be
  dominated by star formation. Others, like MRK~709~S (red circle in
  Figure~\ref{fig:wisecolor}, right), have X-ray and radio emission
  indicative of AGN activity \citep{reines14}. The WISE colours of
  MRK~709~S are $W1-W2=0.323$ and $W2-W3=4.010$, falling
  just below our colour cut. The fact that some of the BCDs do have
  signs of BH accretion activity indicates that the red colours may not
  be solely due to star formation in all cases.

\end{enumerate}

\subsection{Optical line emission diagnostic}
\label{subsec:bptdiag}

We explored the distribution of our sample of 3326 galaxies with WISE
matches and S/N $ > 3$ in $W1$, $W2$ and $W3$ in the diagnostic
diagrams of narrow-line ratios, which are powerful tools for distinguishing
Seyfert galaxies, low-ionisation nuclear emission-line region sources,
and HII galaxies
\citep{baldwin81,veilleux87,ho97,kewley01,kewley06,kauffmann03}. Diagnostic
line-intensity ratios, such as [OIII]/H$\beta$ and [NII]/H$\alpha$,
corrected for reddening, are effective at separating populations with
different ionisation sources. The ionising radiation field found in
active galaxies is harder than in star-forming galaxies, and this
gives higher values of [NII]/H$\alpha$ and [OIII]/H$\beta$. Hence, the
narrow line AGN are found in the upper right portion of the
[OIII]/H$\beta$ versus [NII]/H$\alpha$ diagnostic diagram. The lines
in these ratios are also selected to be close in wavelength space in
order to minimize the effect of dust extinction on the computed line
ratios.

We obtained emission line fluxes and stellar mass estimates for our
sample from the Max Planck Institute for Astrophysics/Johns Hopkins
University (MPH/JHU)
collaboration\footnote{http://www.mpa-garching.mpg.de/SDSS/},
which contains 927552 SDSS galaxies. Of the 968 matches, we found that 954
(9 of the 43 AGN candidates without prior identification) had
all four [OIII], H$\beta$, [NII], and H$\alpha$ line fluxes and we
were therefore able to compute the line-intensity ratios shown in
Figure~\ref{fig:bpt}, left. As in Figure~\ref{fig:wisecolor}, left,
our 135 nearby galaxies are shown as {\it red triangles} and the sample of
126 galaxies, including dwarfs, that had already been identified as
having the optical spectroscopic signature of AGN activity (MRBGD) are
shown as {\it black open triangles}. These additional optically
identified AGN are split into 111 type~1 (BL; {\it green filled
  triangles}) and 16 type~2 (NL; {\it blue filled triangles}). The {\em
  dashed line} is the demarcation line between normal star-forming
galaxies and AGN from \citet{stasinska06}.

\begin{figure*}
\centerline{
\includegraphics[width=240pt,height=240pt,angle=0]{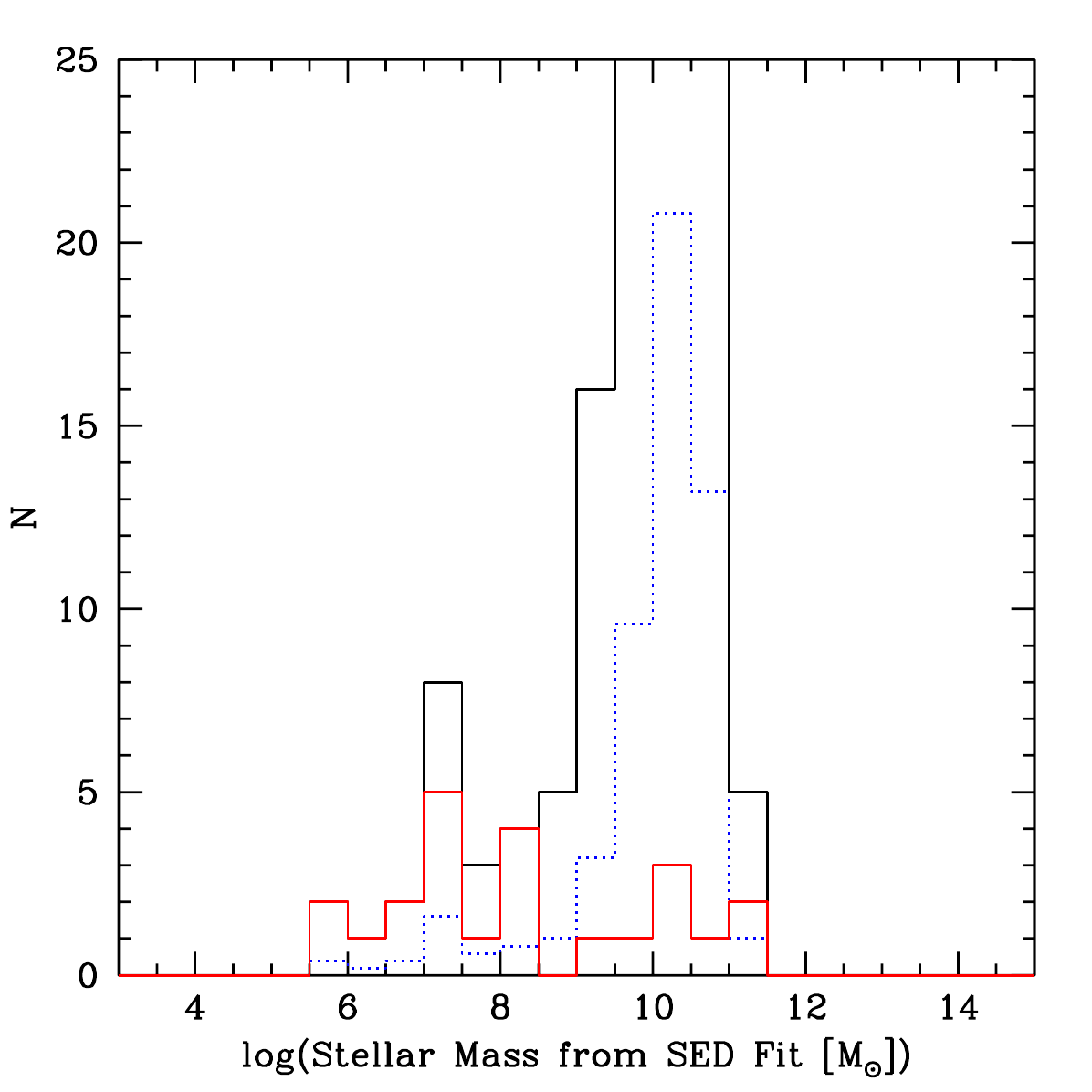}
\includegraphics[width=240pt,height=240pt,angle=0]{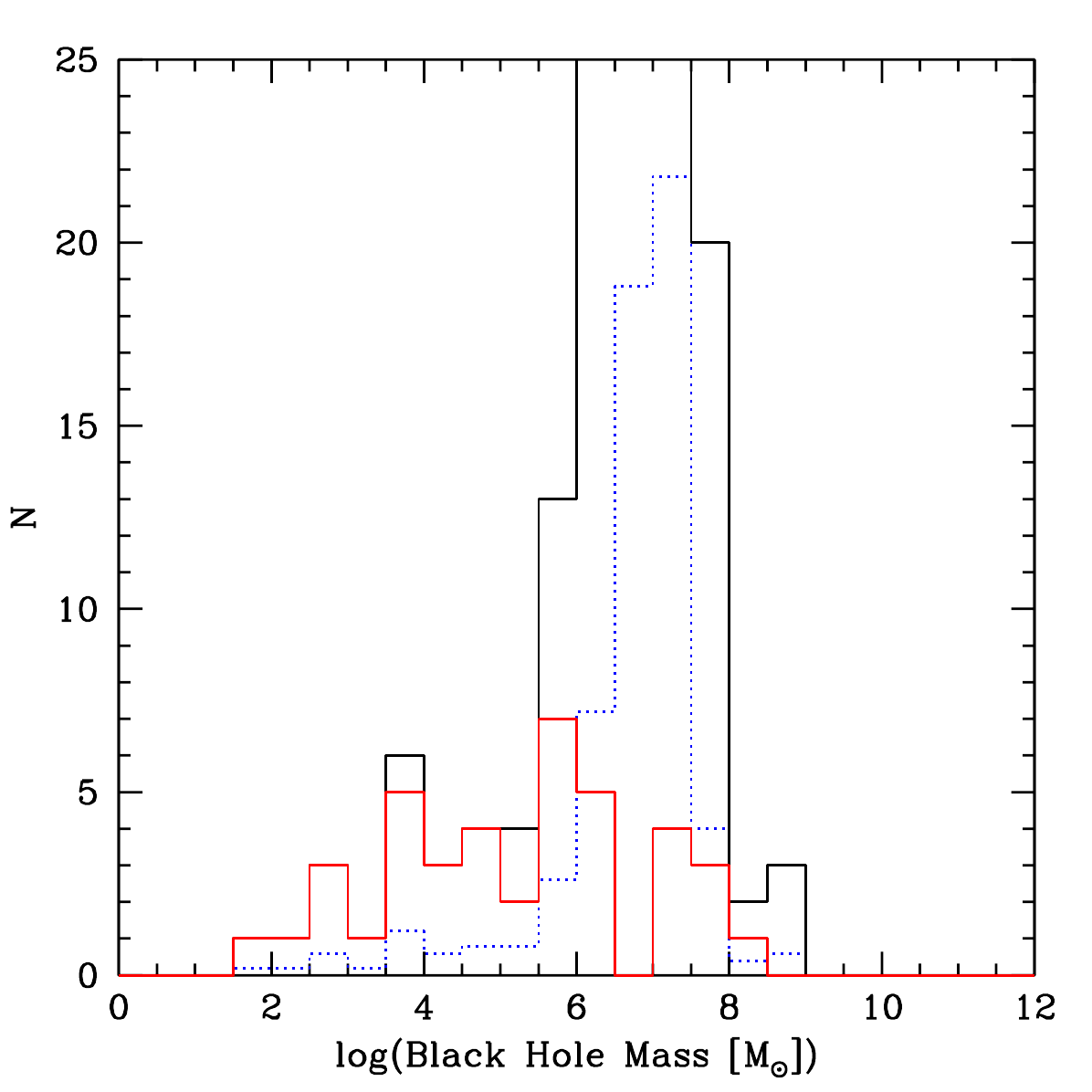}
}
\caption{\label{fig:hist_mstarSED} {\it Left}: SED-based stellar mass
  distribution of our sample of 264 AGN candidates. 
  The subsample of 23 galaxies that are not in the samples of MRBGD,
  shown here in red, contains
  the lowest stellar mass galaxies hosting an AGN candidate known
  today. The SED-based stellar mass distribution of
    our sample of 264 AGN candidates is also shown scaled down by a
    factor of 5 ({\it blue dotted line}). {\it Right}: Black hole mass
  distribution of our sample of 300 AGN candidates with
  estimated black hole mass. The black hole masses
    were estimated from their IR luminosity, assuming $L_{bol} = 0.1
    L_{Edd}$, the mean of our calibration sample. The
  subsample of 40 galaxies selected via the IR diagnostic,
  shown here in red, contains the lowest mass BH candidates known
  today. As in the left plot, the black hole mass
    distribution of our sample of 300 AGN candidates is also shown
    scaled down by a factor of 5 ({\it blue dotted line}). \vspace{0.2cm}}
\end{figure*}

The main results of this comparison is summarized below.

\begin{enumerate}

\item As expected, the majority of the optically classified AGN are
  located in the conventional region of Seyfert galaxies, in terms of
  the semi-empirical demarcation line of \citet{stasinska06} (the {\it
    dashed line} in Figure~\ref{fig:bpt}, left) in the [OIII]/H$\beta$
  versus [NII]/H$\alpha$ diagram. This also includes four objects
  that had not previously been identified as AGN, if we
  conservatively include the two objects close to the demarcation line
  (red triangles in Figure~\ref{fig:bpt}, left).

\item The remaining one-third of the objects (five that had not
  previously been identified as AGN) are located in the region for HII
  galaxies in the same diagram. However, this does not necessarily
  mean that they are not AGN. Indeed, \citet{sartori15} compared AGN
  selected via three methods (using the classical BPT diagram, a
  similar optical emission line diagnostic based on the He II
  4686\AA\ line, and mid-IR colour cuts) and found that only 3 of
  their 336 sources fulfilled all three criteria and that different
  criteria selected host galaxies with different physical properties
  such as stellar mass and optical colour. It is therefore likely that
  our IR colour cut is selecting a different subsample of AGN to that 
  selected by the optical diagnostic diagram. Also, as is reported in
  \citet{dong12}, this difference can be explained by the possible
  inclusion in the SDSS fibre aperture of emission from star formation
  regions in the host galaxies. Hence, although the optical spectra
  undeniably reveal the presence of a broad-line AGN, the narrow-line
  ratios may still be dominated by the characteristic emission of an
  HII region. We also note that MRK~709~S (red circle in
  Figure~\ref{fig:bpt}, right), which has X-ray and radio emission
  indicative of AGN activity \citep{reines14}, is located very close
  to the demarcation line.

\item Our comparison reveals some possible issues regarding using the
  usual demarcation line between normal star-forming galaxies and
  AGN. Indeed, it is possible that the demarcation may not apply to
  the low-mass galaxy regime and for nearby galaxies. Given that the
  BPT diagram is empirically derived, and to date there has been 
  little data on AGN in dwarfs, this is perhaps not surprising. In
  relation to this, it has also been known for some time that star
  formation behaves differently in some dwarf galaxies where there is
  a steepening of the Kennicutt–Schmidt law
  \citep{bigiel08,elmegreen11,roychowdhury15}. Additionally, recent
  work by \citet{kewley15} has shown that at high redshift, the
  demarcation line changes as a function of redshift.  It is therefore
  perhaps possible that such a dependence could also extend to very
  low redshifts and/or low masses.

\end{enumerate}

\section{Distance distribution}
\label{sec:distance}

One of the main goals of this study is to identify the closest dwarf
galaxy hosting an AGN. Hundreds of active massive black hole
candidates have already been detected at the centres of low-mass
galaxies \citep[e.g.][]{marleau13, moran14, reines13}. The major
limitation of these works, however, is that they do not provide an
unambiguous confirmation of the existence of these IMBHs as they rely
solely on detecting the radiative signatures of AGN. Moreover, the
dwarf galaxies in these existing samples are too far away to carry out
dynamical studies \citep[e.g.\ the][sample has a redshift range of $z
  = 0.001 - 0.055$]{reines13}. It follows that in order to
firmly establish that massive black holes do indeed exist at the
centres of dwarf galaxies, it is absolutely necessary to extend our
search to the very nearby Universe, i.e.\ to the galaxies for which
we can resolve the sphere of influence of the CMBH and obtain
dynamical mass estimates.

\begin{figure*}
\centerline{
\includegraphics[width=240pt,height=240pt,angle=0]{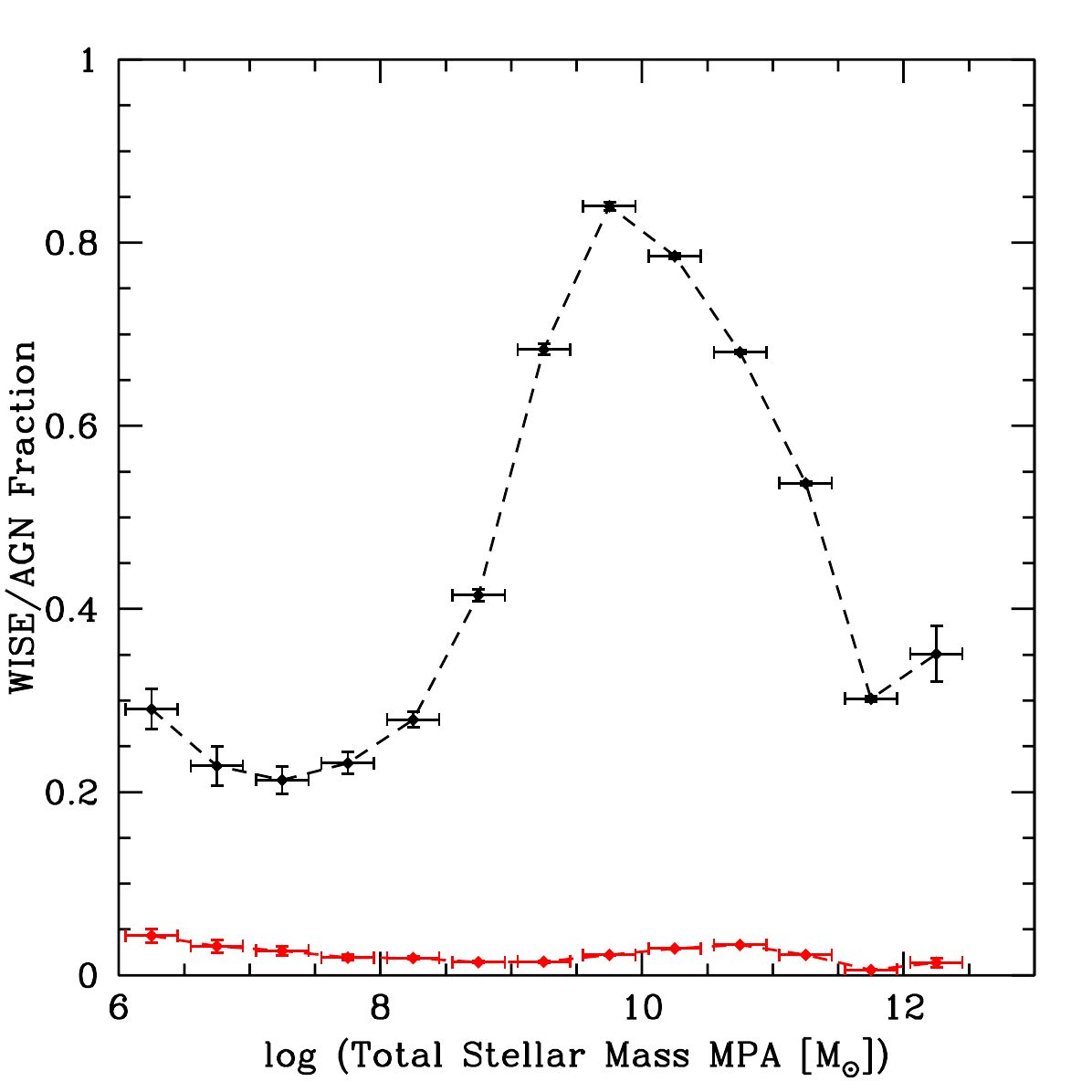}
\includegraphics[width=240pt,height=240pt,angle=0]{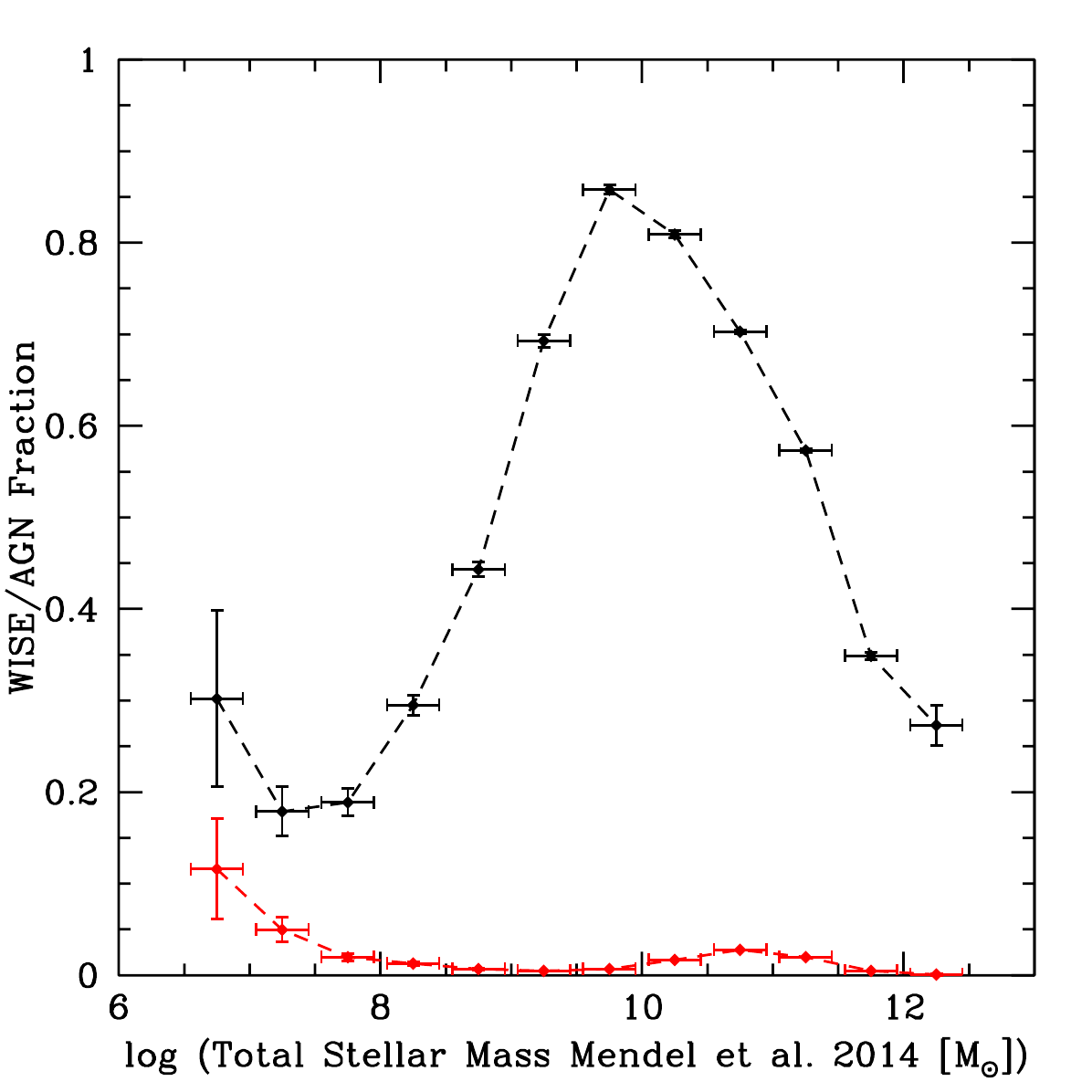}
}
\caption{\label{fig:agnfraction} {\it Left}: WISE ({\it black
    curve})/AGN ({\it red curve}) fraction as a function of stellar
  mass derived from the MPA/JHU catalogue. {\it Right}: Same as left
  but from the catalogue of \citet{mendel14}. \vspace{0.2cm}}
\end{figure*}

In Figure~\ref{fig:hist_dist} we show the distance distribution of 
the 300 out of the 303 galaxies with known redshift/distance. 
The first three candidates listed in
Table~\ref{tbl:tbl-1} have no known redshift/distance. The sample of
40 galaxies selected via the IR diagnostic, shown in red, {\em
  contains the nearest AGN candidates known today}. Below 11 Mpc, we
find 11 AGN candidates, including NGC~4395 (4.5 Mpc) which has
been until now the closest one identified in a dwarf galaxy. Our
method detected AGN in five galaxies that are closer than NGC~4395:
HIPASS~J1247-77 (3.2~Mpc), UGC~04459 (3.2~Mpc), NGC~5253 (3.6~Mpc),
IC~2574 (3.8~Mpc) and CIRCINUS (4.2~Mpc). Of these, four can be
classified as dwarf galaxies based on their stellar mass and/or
absolute magnitude: HIPASS~J1247-77 \citep{ryan-weber02} has a
magnitude $M_B$ = -12.91 \citep{karachentsev04}, UGC~04459 has a
stellar mass of $\sim 9.7 \times 10^{6}$ M$_{\odot}$ and a magnitude
$M_B$ = -13.43 \citep{zhang12}, NGC~5253 \citep{turner15} has a
stellar mass of $\sim 1.5 \times 10^{8}$ M$_{\odot}$ \citep{martin98}
and IC~2574 has stellar mass of $\sim 6.3 \times 10^{7}$ M$_{\odot}$
\citep{lee11} and a magnitude $M_B$ = -16.8 \citep{walter99}. The fifth
galaxy, CIRCINUS, has a stellar mass of $\sim 9.5 \times 10^{10}$
M$_{\odot}$ and is therefore not considered a dwarf galaxy
\citep{for12}. In the context of being able to follow-up and confirm
the presence of a IMBH in a dwarf galaxy with dynamical measurement,
this implies that nearby dwarf galaxies should be targeted for
dynamical observations. 

\section{Stellar mass estimates}
\label{sec:mstar}

We are interested in determining whether low-mass galaxies harbour a CMBH,
and if so, whether or not the correlation between BH mass and total
stellar mass derived in \citet{marleau13} extends to the low-mass
regime. Therefore, in our calculation of stellar and BH masses, we 
did not only consider galaxies in our sample of dwarfs but also included 
the other galaxies in our nearby galaxies sample.

We used a combination of three independent methods to estimate the
stellar masses for the 300 out of 303 galaxies with redshift
measurement. The first method consisted of fitting the SEDs of our
galaxies, constructed from SDSS photometry, using the MAGPHYS package
\citep{dacunha08}. For the second method, we estimated the stellar
masses following the empirical relation of \citet[see their equation
  8]{taylor11}. This method combines a galaxy's luminosity ($M_i$
expressed in the AB system) with a mass-to-light ratio derived from a
colour measurement ($g-i$). We transformed the SDSS magnitudes into AB
mag, applied $K$-corrections using the method of
\citet{chilingarian10}, which, for the low-redshift ($z < 0.01$)
galaxies in our sample, typically affect the $g$ and $i$ values by a
few hundredths of a magnitude. The absolute magnitude $M_i$ was
computed using the distances given in Table~\ref{tbl:tbl-1}. This
method only applies for galaxies with $z < 0.5$ (297 of the 300
galaxies; all higher redshift sources are in the MRBGD samples). The
third method consisted of using the $Ks$-band magnitude taken from
both the PSC and the XSC of 2MASS. A general calibration factor was
applied to the $Ks$ fluxes by comparing the stellar masses to the SED
derived stellar masses.

\begin{figure*}
\centerline{
\includegraphics[width=170pt,height=170pt,angle=0]{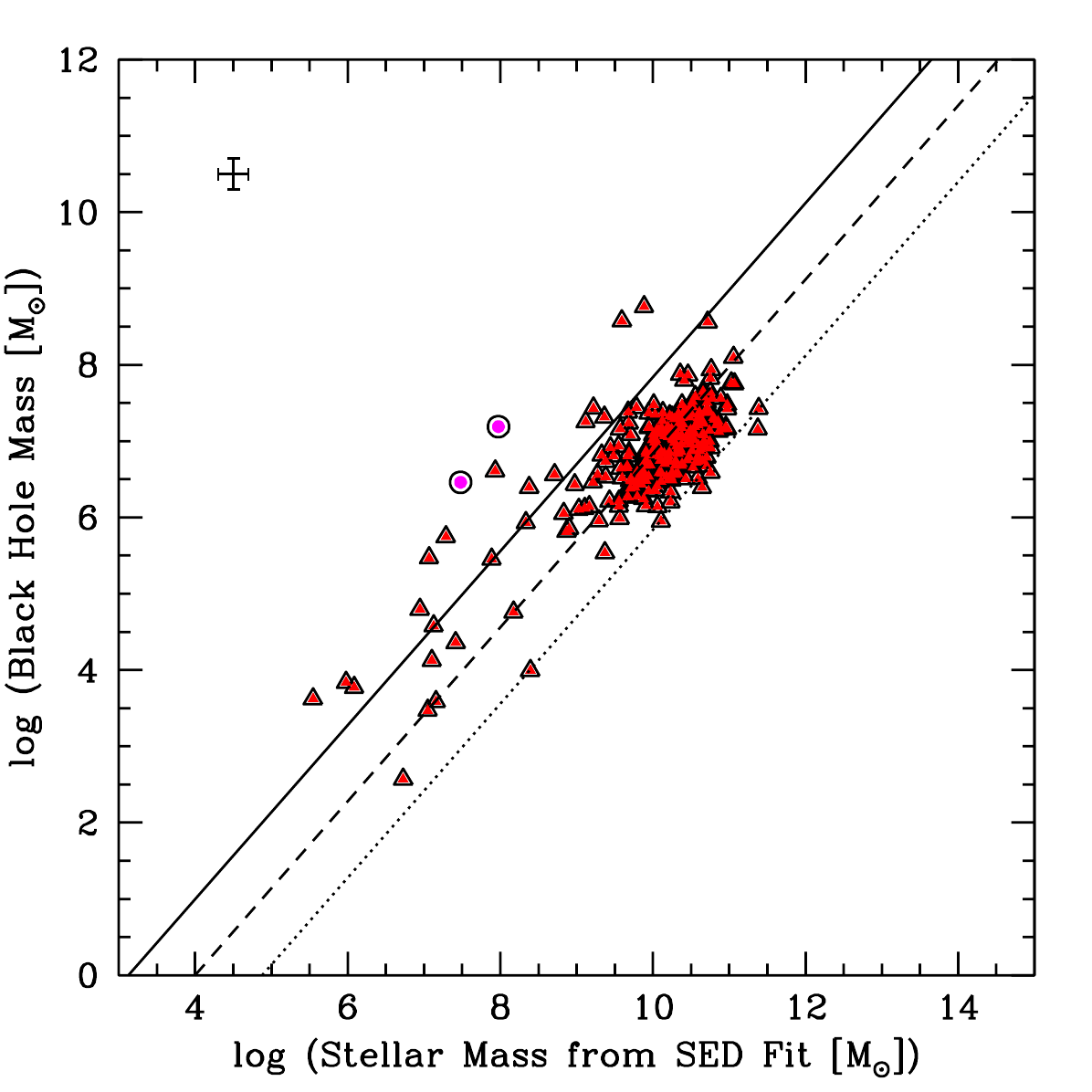}
\includegraphics[width=170pt,height=170pt,angle=0]{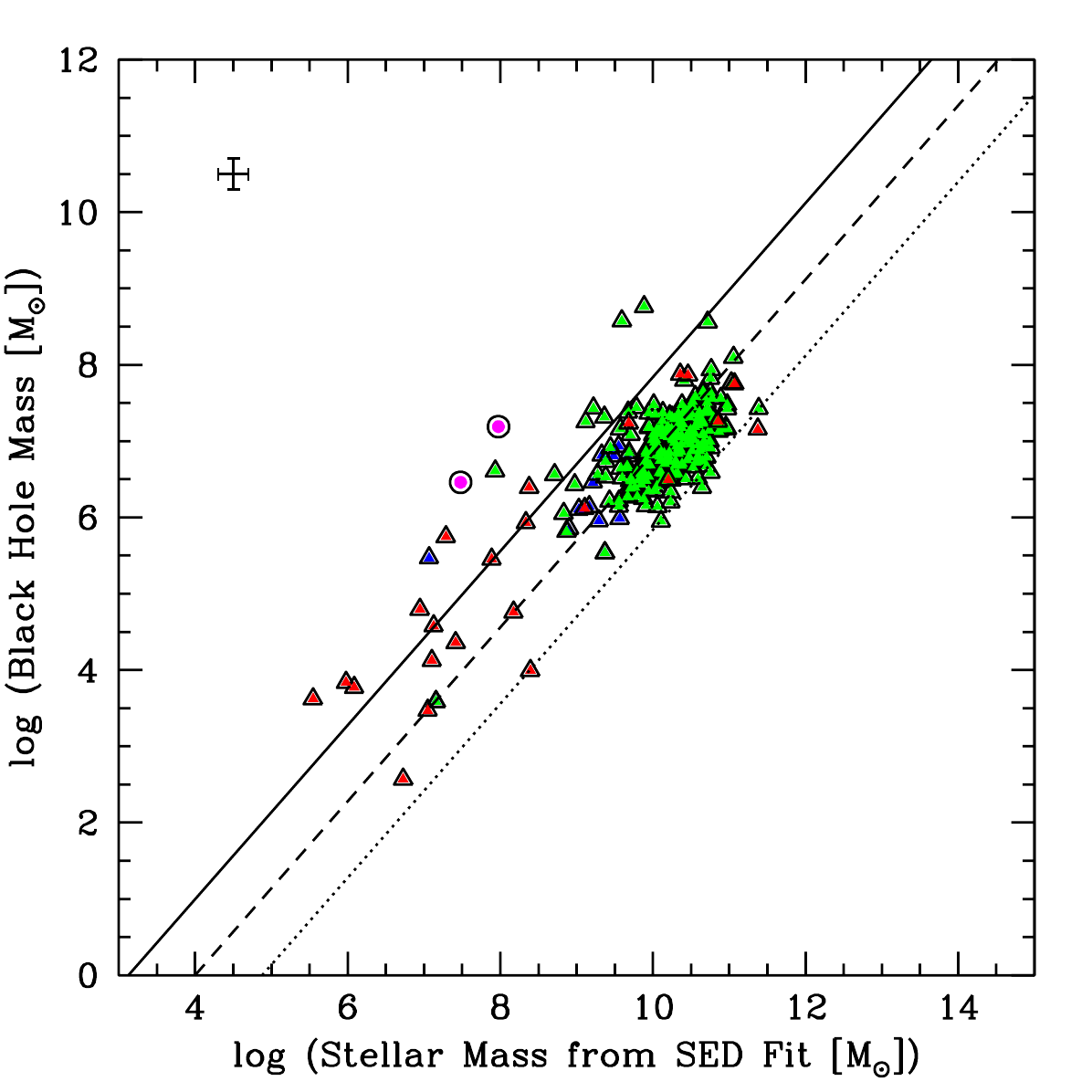}
\includegraphics[width=170pt,height=170pt,angle=0]{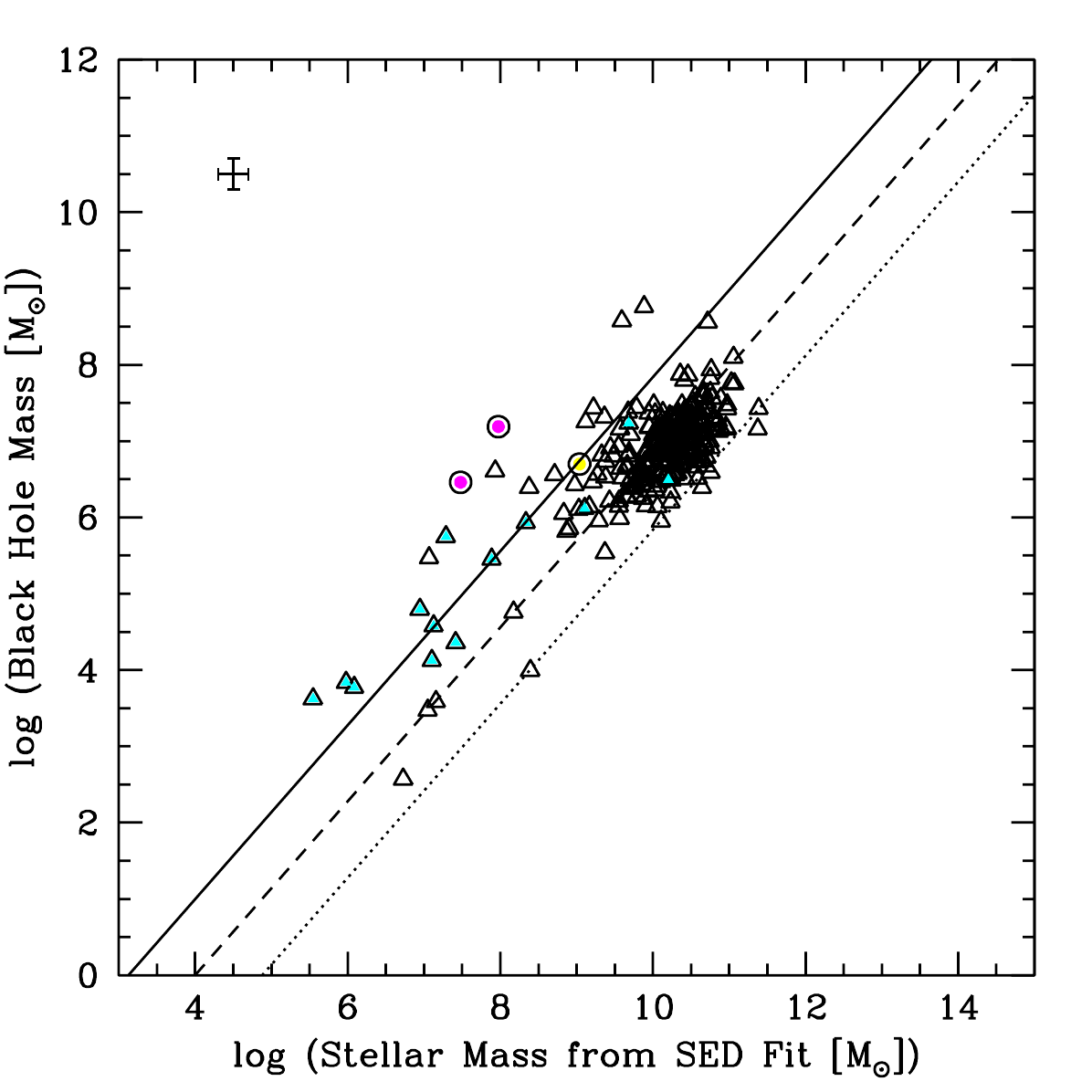}
}
\caption{\label{fig:mstar_bhmass} 
{\it Left}: Total stellar mass from
  SED fit vs.\ black hole mass obtained using the bolometric
  luminosity for our infrared sample of 264 galaxies with both 
  known redshifts/distances and stellar masses 
({\it red filled triangles}). 
The data points and bisector linear regression fit of \citet{marleau13}
  ({\it dashed line}) are plotted for $L_{bol}/L_{Edd} = 0.1$. Also
  shown are the fit for $L_{bol}/L_{Edd} = 1.0$ ({\it dotted line})
  and for $L_{bol}/L_{Edd} = 0.01$ ({\it solid line}). 
{\it Middle}: Same as left, but showing previously identified AGN of 
  type~1 and type-2, respectively shown as {\it green filled triangles}
  and {\it blue filled triangles},
  and the AGN candidates identified from the IR diagnostic only,
  shown as {\it red filled triangles}. Also shown are the two
  low-metallicity and heavily obscured BCDs that were originally
  identified by \citet{griffith11} ({\it magenta filled circles}). 
{\it Right}: Same as left, but showing the BCDs ({\it cyan filled triangles}),  
  the two low-metallicity and heavily obscured BCDs that were
  originally identified by \citet{griffith11} ({\it magenta filled
    circles}), and the BCD MRK~709~S with X-ray and radio
   emission indicative of AGN activity ({\it yellow filled circle}).
    \vspace{0.2cm}}
\end{figure*}

We compared the stellar masses measured from our three methods with
those listed in the MPA/JHU and \citet{mendel14} catalogues and those
found in the literature. For the 261 galaxies with stellar masses from
both the SED fit and the color measurement method, we found in general
very good agreement between the two estimates, with only two
exceptions wherein the two mass estimates differ by an amount larger
than the scatter. We find excellent agreement between our SED fit
stellar masses and those of the MPA/JHU and \citet{mendel14}
catalogues which confirms that our SED fit stellar mass estimates are
robust. We also find in general good agreement between the SED fit
stellar masses and the few stellar masses which were reported by
MRBGD.

As can be seen in Figure~\ref{fig:hist_mstarSED}, we sample evenly a
broad distribution of stellar masses, including dwarf galaxies in the
stellar mass range $\sim 10^6 - 10^9$ M$_{\odot}$.

\section{WISE/AGN fraction}
\label{sec:agnfraction}

As discussed in the Introduction, dwarf galaxies exhibit many
properties that distinguish them from massive galaxies (e.g.\
mass-to-light ratios, star formation histories, metallicities) and
questions exist about the nature of their evolution with respect to
massive galaxies. In this regard, it is interesting to consider
whether AGN have evolved differently in dwarf galaxies than in massive
galaxies. A measure of this is given by the AGN fraction at a given
time and mass.

Starting from the MPA/JHU catalogue of galaxies with stellar masses
and from the catalogue of \citet{mendel14}, we examine the fraction of
WISE matches and IR selected AGN as a function of stellar mass in each
of these samples. Each catalogue was cross-matched independently to
the WISE All-Sky Release Source Catalog using the same method as
described in Section~\ref{sec:IRcolor} and the same colour cut as described in
Section~\ref{sec:IRAGN}. The errors were computed assuming Poisson
statistics. As can be seen in Figure~\ref{fig:agnfraction}, the
fraction of IR selected AGN shows a signature bump at a stellar mass
$\sim 5.6 \times 10^{10}$ M$_{\odot}$. Also, the fraction of AGN
appears to increase as a function of decreasing stellar mass at
stellar masses below $\sim 10^{9}$ M$_{\odot}$, i.e.\ in the low-mass
regime of dwarf galaxies, an effect also reported in \citet[][see
  their Figure 5]{satyapal14}. Note, however, as can also be seen in
the figure, that this increase is accompanied by an increase in the WISE
fraction. This is not the case for the bump seen at $\sim 5.6 \times
10^{10}$ M$_{\odot}$, which is displaced from the peak in the WISE
fraction. If real, this behaviour at low mass could be due to 1) the
fact that BH accretion activity is higher in nearby dwarf galaxies
than in their more massive counterparts (similar to the star-forming
activity, i.e.\ the ``downsizing'' effect); 2) the fact that it is
easier to detect WISE sources/AGN in nearby low-mass (low surface
brightness) galaxies; or 3) the fact that at these low masses, the 
fraction of AGN candidates contaminated by star formation may 
be higher. This is a result that should be further explored and 
confirmed.

\begin{figure}
\centerline{
\includegraphics[width=240pt,height=240pt,angle=0]{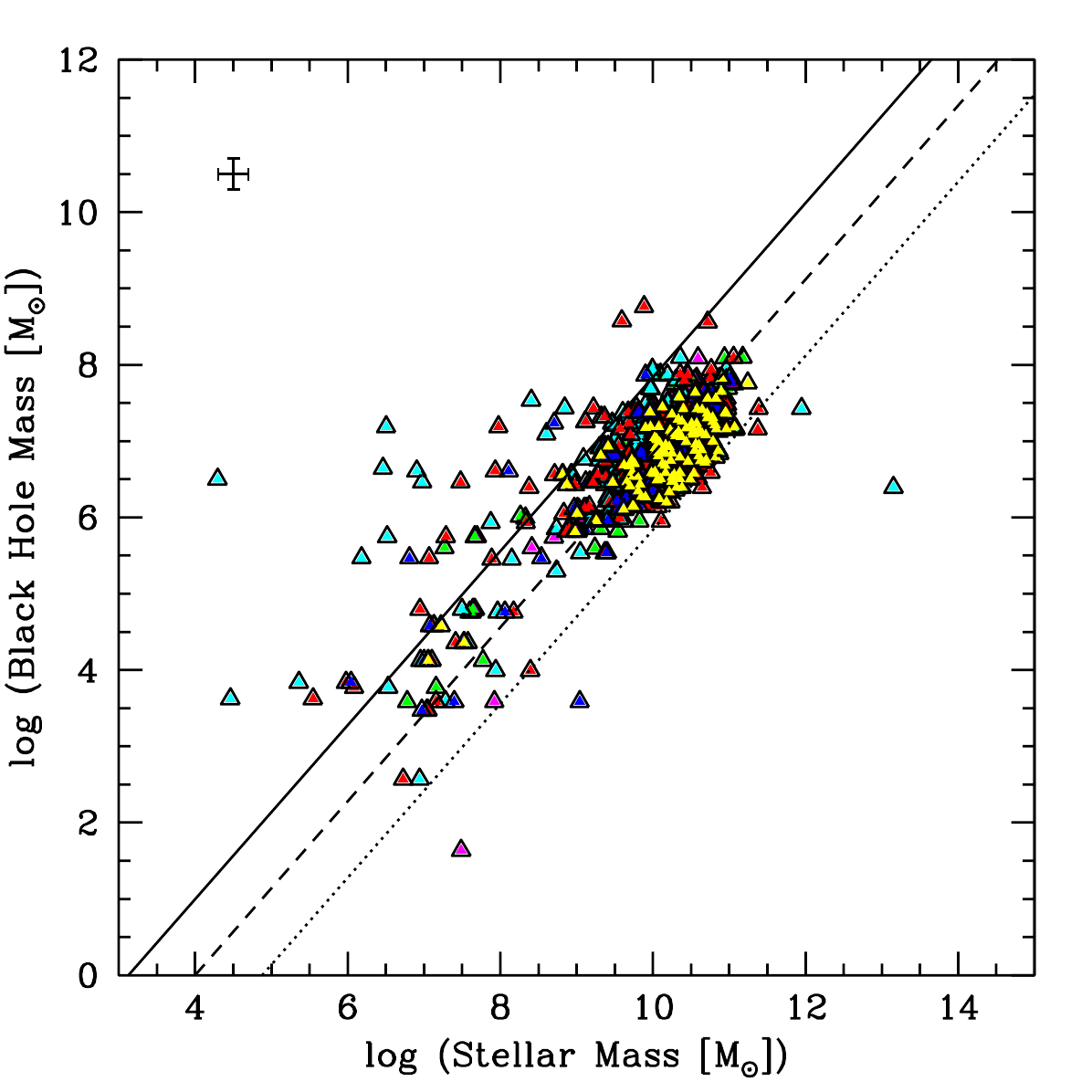}
}
\caption{\label{fig:mstarall_bhmass} Total stellar mass from all methods used for
  stellar mass measurements vs.\ black hole mass obtained using the
  bolometric luminosity for our infrared sample of 300 galaxies: from
  SED fit ({\it red filled triangles}), from colour measurement ({\it
    cyan filled triangles}), from $Ks$-band from 2MASS/PSC and XSC
  ({\it magenta filled triangles}), from the literature ({\it blue filled
    triangles}), from \citet{mendel14} ({\it yellow filled triangles})
  and MPA/JHU catalogues ({\it green filled
    triangles}). \vspace{0.2cm}}
\end{figure}

\section{Black hole mass estimates}
\label{sec:bhmass}

Black hole masses were estimated using the bolometric luminosity of
our AGN candidates. Bolometric luminosities, taken to be the 100
$\mu$m to 10 keV integrated luminosity \citep{richards06}, are
typically obtained using corrections to the mid-infrared bands where
the AGN emission dominates. The 12 (W3) and 22 $\mu$m (W4) k-corrected
flux densities from WISE can be used to compute the bolometric
luminosities of the WISE colour-selected AGN by applying the bolometric
corrections $L_{bol} \simeq 8 \times L_{12\mu {\rm m}}$ (W3) and $L_{bol}
\simeq 10 \times L_{22\mu {\rm m}}$ (W4) from \citet{richards06}, which are
not strongly dependent on AGN luminosity (see their
Figure~12). However, note that \citet{richards06} only considered
type~1 (BL) AGN in their analysis so that applying these corrections
to type~2 AGN may lead to additional uncertainties in the BH mass
estimates. Here, we only compute the BH mass estimates based on the 12
$\mu$m luminosities, as the S/N of the WISE data in the 12 $\mu$m 
images is higher than in the 22 $\mu$m images. By comparison, the
bolometric luminosity of an AGN can also be estimated using the [O
  III]~$\lambda$5007 emission line luminosity
\citep[e.g.][]{moran14}. In this case, the bolometric correction
ranges from a factor of a hundred ($L_{bol} \simeq 3500 \times
L_{5007}$) \citep{heckman04} to a factor of ten ($L_{bol} \simeq 174
\times L_{5007}$) \citep{greene07} larger than the IR bolometric
correction.

For point source AGN candidates, we used the w3mpro magnitudes to
compute BH masses. For our 30 extended source AGN candidates, we
used the aperture magnitude corresponding to the measured extent of
the central source in W3 (e.g.\ for a galaxy with a radial
extent of 19 arcsec in W3, we used w3mag\_6). As 11 of our AGN
candidates are spatially larger than the largest WISE aperture (24.75
arcsec radius aperture), the BH mass estimates for these sources is
only a lower limit. Given the bolometric luminosity derived from the
IR, making the assumption that accretion is at the Eddington limit
yields a lower limit on the black hole mass. However, given that the
growth rates of AGN can span several orders of magnitude, from
super-Eddington accretion to $10^{-3} L_{Edd}$ \citep{simmons13,
  steinhardt10}, we computed black hole masses assuming the following
three cases: $L_{bol} = L_{Edd}$, $L_{bol} = 0.1 L_{Edd}$ and $L_{bol}
= 0.01 L_{Edd}$.

In a separate work \citep{lopez15}, the IR CMBH mass estimates were calibrated 
using 113 galaxies with robust CMBH mass measurements. Of the 113 galaxies, 51
have estimates based on the reverberation mapping method and 62 based
on dynamical measurements. After following the procedure described in
Sections~\ref{sec:IRcolor} and \ref{sec:IRAGN}, 50 of these galaxies
were identified as AGN candidates based on our IR colour diagnostic and
their bolometric luminosity was computed as described in the
paragraphs above. By comparing the CMBH mass estimates based on the
other two methods to the estimate based on the IR method, the distribution
of $L_{bol}/L_{Edd}$ was computed. A Gaussian fit to this
distribution, in log space, yielded a mean of -1.0 and a $\sigma$ of
0.6, in agreement with the range of Eddington ratios found in the
literature \citep{woo02, mushotzky08}.

The histogram of BH masses computed using $L_{bol} = 0.1 L_{Edd}$, the
mean value derived from our calibration sample (-1.0 in log space), is
shown in Figure~\ref{fig:hist_mstarSED}, right. This choice of
$L_{bol}/L_{Edd}$ was also used in \citet{marleau13} and also appears 
to be a good fit to this data set (see
Section~\ref{sec:relation}). The black hole masses associated with the
dwarf galaxies range from $\sim 10^3 - 10^6$ M$_{\odot}$ (see 
Section~\ref{sec:relation}), which is a significantly lower range than
has previously been probed. However, as stated above, note that the
Eddington ratio is known to vary by as much as three orders of
magnitudes for different AGN so our results are also presented for
values of $L_{bol}/L_{Edd}$ equal to 0.01 and 1.0.

For NGC~4395, since an Eddington ratio has been measured with a value of
0.0012 \citep{peterson05}, we can compute its IR CMBH mass directly.
We calculate a mass of $3.6 \times 10^5$ M$_{\odot}$, in very good
agreement with the dynamical mass of $4 \times 10^5$ M$_{\odot}$
given by \citet{denBrok15}. POX~52 also has a reported Eddington
ratio of $0.2 - 0.5$ \citep{thornton08}, which we can use to compute
an IR CMBH mass. Using a ratio of 0.2, we obtain an IR CMBH mass of 
$2.0 \times 10^5$ M$_{\odot}$, in agreement with the range of CMBH 
masses of $2.2 - 4.2 \times 10^5$ M$_{\odot}$ obtained by \citet{thornton08} 
using other CMBH mass estimator methods.

\begin{figure}
\centerline{
\includegraphics[width=240pt,height=240pt,angle=0]{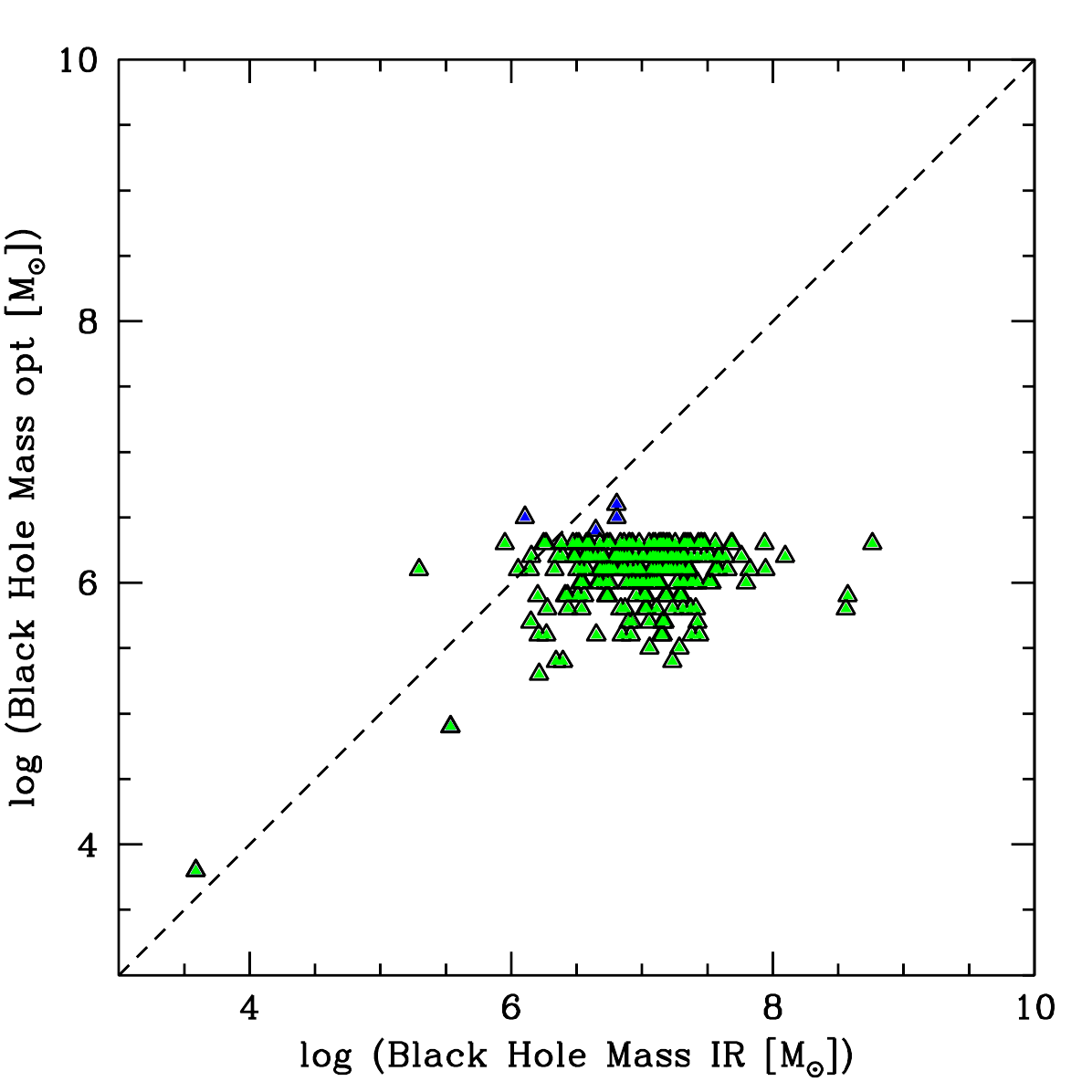}
}
\caption{\label{fig:bhmassIR_bhmassopt} Comparison between the black
  hole mass estimates computed from their IR luminosity and the black
  hole mass computed from the [OIII]~$\lambda$5007 line flux from the
  works of MRBGD, for a subsample of 248 galaxies for which we were
  able to find a match ({\it black open triangles}). Previously optically
  identified type~1 and type~2 AGN are shown as {\it green filled triangles}
  and {\it blue filled triangles}, respectively. \vspace{0.2cm}}
\end{figure}

\section{BH mass versus stellar mass scaling relation}
\label{sec:relation}

Using the stellar masses of the galaxies and the black hole masses
estimated from the bolometric luminosity of the AGN candidate, we
present in Figure~\ref{fig:mstar_bhmass}, left, the correlation
between black hole mass and total SED stellar mass for our sample of
264 galaxies. We find active BH in nearby dwarfs, as well as {\it in
  dwarf galaxies with lower stellar masses ($\sim 10^6 - 10^9$
  M$_{\odot}$) and correspondingly lower BH masses ($\sim 10^3 - 10^6$
  M$_{\odot}$)} than the previous works of MRBGD, which only probed to
$10^8-10^9$ stellar masses. We also find that {\it the current
  results are consistent with the existing correlation
  \citep{marleau13} extending linearly (in log-log space) into the
  lower mass regime}. However, we cannot rule out with the current
data that the correlation could be weakly non-linear in the low-mass
regime.

In Figure~\ref{fig:mstar_bhmass}, middle, we examine the stellar mass
versus BH mass relationship of the BCDs. We see that in the low-mass
regime, the BCDs tend to be located mostly above the relation. This
may indicate that a possible upturn at low mass is not real but due to
contamination from star formation activity. Indeed, we find that the
actively star-forming BCDs of \citet{griffith11} ({\it magenta
  circles}) fall above the relation. Also, the AGN hosting BCD
MRK~709~S \citep{reines14} ({\it red circle}) is expected to be
affected by the host galaxy light based on its IR colour ($W1-W2 =
0.323$). Its position in the diagram does indeed show a value slightly
above the relation.

The errors on the stellar masses were determined using the
uncertainties associated with the SED fit and were found to be of the
order of 0.2~dex (97.5 percentile or 2$\sigma$ error). In
Figure~\ref{fig:mstarall_bhmass}, we explore the effect of the
uncertainty introduced by using different methods on the scatter of
the stellar mass versus BH mass relation. The different colours are
associated with the different methods and catalogues used in
estimating the stellar masses. Except for the colour measurement method, 
we find that the agreement between the methods is quite good
(see also Section~\ref{sec:mstar}) and does not significantly change
the scatter seen in Figure~\ref{fig:mstar_bhmass}, left.

The errors on the BH mass were calculated from the uncertainties in
the bolometric luminosity measurements and were found to be
$\sim$0.2~dex. However, the main source of uncertainty in this
measurement is the value of $L_{bol}/L_{Edd}$, which we have assumed to
be the same for all sources but instead most likely varies for each
source. As can be seen in Figure~\ref{fig:mstar_bhmass}, simply
allowing this value to vary from 0.01 to 1.0 reproduces most of the
scatter seen in the data.

We compare in Figure~\ref{fig:bhmassIR_bhmassopt} the black hole mass
estimates computed from the IR luminosity and those computed from the
[OIII]~$\lambda$5007 line flux from the works of MRBGD, for the
subsample of 248 galaxies for which we were able to find a match. The
BH mass cut-off of \citet{greene07} was $2 \times 10^6$\msol\ (6.3 in
log space, see their Figure 1) and this cut-off is easily seen in
the diagram. Although some BH mass estimates are in agreement, the
majority are not. The BH mass estimates computed from the
[OIII]~$\lambda$5007 line flux are systematically lower than the BH
mass estimates computed using the IR method.

\section{Discussion and conclusions}
\label{sec:discussion}

We undertook a census of central massive black holes in the very
nearby Universe, targeting specifically low-mass dwarf systems with BH
masses in the IMBH mass range. Additionally, we were interested in
detecting AGN in nearby galaxies regardless of their mass so that we
could identify targets for future dynamical mass measurement.  The
results of our paper are as follows:

\begin{enumerate}

\item Using the WISE All-Sky Release Source Catalog, we examined the
  IR colours of a sample of known low-mass and other nearby systems in
  order to identify candidate AGN by applying the infrared colour
  diagnostic $W1-W2 > 0.5$. We find that 303 nearby galaxies have WISE
  colours consistent with galaxies containing an AGN.

\item We validate our IR detection method. Of the 303 candidate AGN, 
  276 (or 91\%) are subsequently found to have been independently 
  identified as AGN via other methods. The remaining 9\% require 
  follow-up observations  
  to confirm that their red IR colours ($W1-W2 > 0.5$) are not due to 
  star formation. We find that if we applied an additional cut in
  $W2-W3$ colour, e.g.\ $W2-W3 < 4.2$, we would reject 16 AGN
  candidates, 8 of which are optically identified AGN (2 type 1 and 6
  type 2) and only 2 are known low-metallicity and heavily obscured
  BCDs. The remaining 6 have colours similar to these 8 optically
  identified AGN and hence are valid candidate AGN. We also point out
  that AGN detected via other methods in dwarf galaxies, such as the
  dwarf galaxy Henize 2-10, have $W2-W3$ colours $> 4.2$ ($W2-W3 \sim
  5.0$) and do not fall within the \citet{jarrett11} demarcation.

\item We verify our IR detection of AGN candidates by comparing the data with
  existing catalogues and optical line emission diagnostics. We find
  that type~1 AGN appear on average to have redder $W1-W2$ IR colours
  and lie preferentially above the IR cut-off, as compared to the
  type~2 AGN which lie preferentially below. We show that NGC~4395,
  POX~52 and UM~625 have WISE colours above our selection cut-off. We
  point out that although some low-metallicity and extremely obscured BCDs
  have very red colours, others are not so red, such as MRK~709~S with 
  $W1-W2=0.323$ and $W2-W3=4.010$.

\item We find that 11 of our candidate AGN lie within a distance
  of 11 Mpc and are the nearest AGN candidates known today. Based on
  our stellar mass cut, we also find that our AGN candidate sample
  contains 62 dwarf galaxies.

\item Using a combination of three independent methods, we obtain
  stellar masses for the galaxies in our AGN candidate sample in the
  range $\sim 10^6 - 10^{11}$ M$_{\odot}$, extending far beyond the
  dwarf galaxy mass demarcation.

\item We show that the fraction of IR selected AGN shows a signature
  bump at a stellar mass $\sim 5.6 \times 10^{10}$ M$_{\odot}$. Also,
  the fraction of AGN appears to increase as a function of decreasing
  stellar mass at stellar masses below $\sim 10^{9}$ M$_{\odot}$,
  i.e.\ in the low-mass regime of dwarf galaxies.

\item Black hole masses are estimated using the bolometric luminosity
  of the AGN candidates and computed for three cases of
  bolometric-to-Eddington luminosity ratio. Using the previously
  measured Eddington ratio of 0.0012, we calculate for NGC~4395 an IR
  CMBH mass of $3.6 \times 10^5$ M$_{\odot}$, in agreement with the
  recently measured dynamical mass estimate of $4 \times 10^5$
  M$_{\odot}$. Similarly, using a previously measured Eddinton ratio of $0.2$, 
  we compute for POX~52 an IR CMBH mass of $2.0 \times 10^5$ M$_{\odot}$, 
  in agreement with the mass range of $2.2 - 4.2 \times 10^5$ M$_{\odot}$ 
  obtained using other methods.

\item Assuming all of our candidates are AGN, we find that activity is
  detected in dwarf galaxies with stellar masses from approximately $10^6$
  to $10^9$ M$_{\odot}$ and that this activity is due to black holes
  with masses in the range $\sim 10^3-10^6$ M$_{\odot}$, assuming
  $L_{bol} = 0.1 L_{Edd}$, the mean of our calibration sample. The
  black hole masses probed here are several orders of magnitude
  smaller than previously reported for centrally located massive black
  holes.

\item We examine the stellar mass versus black hole mass relationship
  in this low galaxy mass regime. The current results are consistent
  with the existing correlation extending linearly (in log-log space)
  into the lower mass regime. However, we cannot rule out with the
  current data that the correlation could be weakly non-linear in the
  low-mass regime.

\end{enumerate}

These results suggest that central massive black holes are present in
low-mass galaxies and in the Local Universe, and provide new impetus
for follow-up dynamical studies of quiescent black holes in local
dwarf galaxies.

\begin{acknowledgements}
This work is a revised version of \citet{marleau14}.  This research
has made use of the NASA/IPAC Extragalactic Database which is operated
by the Jet Propulsion Laboratory, California Institute of Technology,
under contract with the National Aeronautics and Space
Administration. This research has made use of the SIMBAD database,
operated at CDS, Strasbourg, France.
\end{acknowledgements}

\end{document}